\documentclass[journal, doublecolumn]{IEEEtran}
\usepackage{amssymb}
\usepackage{amsthm}
\usepackage{amsmath}
\usepackage{graphicx}
\usepackage{algorithm}
\usepackage{algorithmic}
\usepackage{cite}
\usepackage{bm}
\usepackage{float}
\usepackage{multirow}
\usepackage{color}
\usepackage{subfigure}
\usepackage{caption}
\usepackage{epstopdf}
\usepackage{hyperref}
\usepackage{fixltx2e}
\usepackage{longtable}
\usepackage{diagbox}
\usepackage{xcolor}
\usepackage{changepage}
\usepackage{amsmath, amsfonts}
\usepackage{dsfont}
\usepackage{makecell}
\usepackage{optidef}
\usepackage{stfloats}

\theoremstyle{definition}

\theoremstyle{remark}
\newtheorem{remark}{Remark}
\theoremstyle{plain}
\newtheorem{theorem}{Theorem}
\newtheorem{lemma}{Lemma}
\newtheorem{proposition}{Proposition}
\newtheorem{corollary}{Corollary}

\begin{document}
\title{Performance Analysis of Discrete-Activation Pinching-Antenna Systems}

\author{Yanyu~Cheng,~\IEEEmembership{Senior Member,~IEEE}, Yantao~Pan, Kah Chan Teh,~\IEEEmembership{Senior Member,~IEEE}, Xing~Jin, and~Symeon~Chatzinotas,~\IEEEmembership{Fellow,~IEEE}

\thanks{Yanyu~Cheng and Kah Chan Teh are with the School of Electrical and Electronic Engineering, Nanyang Technological University, Singapore 639798 (e-mail: yanyucheng@ntu.edu.sg; ekcteh@ntu.edu.sg).}
\thanks{Yantao~Pan and Xing~Jin are with the School of Cyberspace, Hangzhou Dianzi University, Hangzhou 310018, China (e-mail: 252270076@hdu.edu.cn; jinxing@hdu.edu.cn).}
\thanks{Symeon~Chatzinotas is with the Interdisciplinary Centre for Security, Reliability and Trust (SnT), University of Luxembourg, Luxembourg L-1855, Luxembourg (e-mail: symeon.chatzinotas@uni.lu).}
}
\maketitle

\begin{abstract}
This paper investigates discrete-activation pinching antenna (DAPA) systems, where a finite set of pinching antennas (PAs) is deployed along a dielectric waveguide, from which PAs are selectively activated to serve users. Performance analyses are developed for both orthogonal multiple access (OMA) and non-orthogonal multiple access (NOMA) by explicitly accounting for the system geometry and discrete PA activation. Specifically, analytical expressions for the outage probability (OP) and ergodic rate (ER) are derived. Furthermore, the discrete-to-continuous gain ratio (DCGR) is introduced to quantify the normalized ER gain of a finite PA deployment relative to the single-PA and continuous-activation references. Monte Carlo simulations validate the derived expressions and show that DAPA systems can closely approach continuous-activation performance with only a moderate number of candidate PAs. For example, under the considered setup, DAPA with five PAs recovers approximately $92\%$ and $90\%$ of the continuous-activation ER gain under OMA and NOMA, respectively, at a transmit power of $0$~dBm.

\end{abstract}


\IEEEpeerreviewmaketitle

\section{Introduction}\label{Sec_introduction}

As wireless communications evolve toward the 6G era, future networks are expected to exploit increasingly higher-frequency spectrum resources while supporting massive connectivity and stringent quality-of-service requirements~\cite{Saad2020Vision6G,Zhang2019Vision6G,Dang2020What6G}. Millimeter-wave and terahertz bands provide abundant spectrum resources but face severe path loss, vulnerability to blockage, and rapidly varying channel conditions~\cite{Andrews2014What5G,Rappaport2015mmWaveChannel,Rappaport2019Wireless,Han2015THz}. These propagation challenges limit conventional fixed-antenna transceivers, whose static radiation structures cannot adapt to dynamic wireless environments~\cite{Heath2016mmWaveMIMO}. Transceiver-side spatial reconfiguration can therefore improve channel conditions without relying solely on additional RF chains or dense antenna deployments.

Representative flexible antenna architectures include fluid antenna (FA) and movable antenna (MA) systems, which exploit spatial diversity by repositioning the effective radiation port within a predefined region~\cite{Wong2021FAS,Zhu2024MA}. This spatial flexibility has also been applied to interference suppression through FA multiple access~\cite{Wong2022FAMA}. Nevertheless, the displacement ranges of these systems are constrained by antenna dimensions and fluidic or mechanical implementations, while mechanically movable structures may incur additional control complexity and energy consumption. Such locally constrained repositioning may be insufficient to compensate for the severe large-scale propagation loss encountered in high-frequency communication systems.

To achieve spatial reconfiguration over a larger range, pinching-antenna (PA) systems have recently emerged as a promising architecture for high-frequency wireless communications. In this architecture, a dielectric waveguide serves as both a transmission medium and a reconfigurable radiation structure. The waveguide supports low-loss surface-wave propagation, while multiple PAs can be deployed or activated along it according to user locations~\cite{Liu2025PASS,Yang2025PAReview,Ouyang2025ArrayGain}. Unlike FA and MA systems, whose effective radiation ports are reconfigured within a prescribed local region, PA systems realize spatial reconfiguration by forming radiation points at different locations along a dielectric waveguide~\cite{Yang2025PAReview,Wang2025PASSBeamforming}. Mechanically repositioned MA implementations rely on actuators or motors and may incur additional movement delay and actuation-energy consumption~\cite{Ding2026MAEnergy,Li2026MATrajectory}, whereas PA systems can provide a larger spatial reconfiguration range without physically translating the radiating element over that range. This spatial flexibility can reduce PA--user distances and improve propagation conditions, while its practical implementation overhead depends on the PA positioning or activation mechanism.

Depending on the activation mechanism, PA systems can employ either continuous or discrete activation. Continuous activation provides fine-grained spatial reconfiguration but may require more stringent position control, hardware support, and signaling overhead. In contrast, discrete activation selects from a finite set of pre-deployed candidate PAs through electronic switching, thereby simplifying practical implementation~\cite{Liu2025PASS}. However, the resulting spatial quantization alters the PA--user distance statistics and consequently affects system performance.

\subsection{Related Work}

Recent advancements in reconfigurable antenna technologies have progressed along several interrelated directions. In the following, we review the most relevant studies that provide the foundation for the proposed DAPA framework.

\subsubsection{Flexible and Fluid Antennas}

Flexible and fluid antennas have attracted significant attention because of their ability to adapt antenna configurations to changing wireless environments. Zheng et al.~\cite{Zheng2024FlexiblePosition} provided a comprehensive overview of flexible-position antenna architectures and their potential applications in future wireless networks. To further exploit spatial adaptability, Wong \emph{et al.} proposed fluid antenna (FA) systems in~\cite{Wong2021FAS}, where a fluid antenna switches among multiple positions within a confined region to select favorable channel conditions. Their theoretical analysis showed that an FA system can outperform conventional antenna-selection combining with multiple fixed antennas while requiring only a single radio-frequency chain.

By adjusting the effective radiation position, flexible and fluid antennas provide additional spatial degrees of freedom compared with conventional fixed-position antennas. However, existing studies on these architectures mainly focus on mitigating small-scale fading and improving diversity performance in non-line-of-sight (NLoS) environments. Their locally constrained reconfiguration differs from the larger-scale geometric adaptation considered in PA systems.

\subsubsection{Pinching Antenna Systems}

Recent studies have established the fundamental architecture and operating principles of PA systems. In~\cite{Liu2025PASS}, the authors presented the basic architecture of PASS and introduced pinching beamforming, in which PAs are attached to dielectric waveguides and positioned at favorable locations. Their results showed that the spatial reconfigurability of PAs can facilitate strong line-of-sight (LoS) links and reduce propagation loss compared with conventional fixed-antenna systems. Yang \emph{et al.} subsequently provided a comprehensive overview of the operating principles, communication architectures, and potential applications of PA systems in~\cite{Yang2025PAReview}, including their integration with non-orthogonal multiple access (NOMA), integrated sensing and communication, and other emerging wireless technologies. The performance of PASS has also been compared with that of reconfigurable intelligent surfaces (RIS) in mmWave systems~\cite{Samy2025PASSRIS}.
More recently, segmented waveguide-enabled PA architectures have been investigated for integrated sensing and communication, where the PA positions and beamforming variables are jointly optimized to improve sensing performance under communication constraints \cite{Geng2026CRLB}.

The multiuser capability of PASS has further motivated its integration with multiple-access techniques. The multiplexing architecture proposed in~\cite{Liu2025PASS} demonstrated that multiple users can be simultaneously served through joint baseband processing and pinching beamforming, with reconfigurable PA positions providing an additional spatial design dimension. The integration of PASS with advanced multiple-access schemes, particularly NOMA, was also discussed in~\cite{Yang2025PAReview}. Subsequent studies investigated NOMA-assisted PA systems from optimization perspectives, including antenna activation, downlink sum-rate maximization, QoS-aware power and position design, and uplink sum-rate or energy-efficiency optimization~\cite{Wang2025NOMAPASSActivation,Zhou2025NOMAPASSSumRate,Xu2025QoSNOMAPASS,Zeng2025UplinkNOMAPASS,Zeng2025EENOMAPASS}, and analytical comparisons between OMA and NOMA~\cite{Cheng2025PASSOMANOMA,Cheng2026SingleWaveguide}. Comparisons among NOMA, frequency-division multiple access (FDMA), and time-division multiple access (TDMA) further showed that the preferred multiple-access strategy depends on user rate requirements and PA-placement flexibility~\cite{Ren2025PASSMeetMultipleAccess}.

Despite these developments, the impact of practical discrete PA activation has not been sufficiently investigated. In particular, the effects of a finite number of candidate PAs on outage probability (OP), ergodic rate (ER), and spatial-gain efficiency under different multiple-access schemes have not been systematically characterized.

\subsubsection{Discrete versus Continuous Antenna Architectures}

The practical performance of PASS depends not only on its ability to reconfigure the effective radiation position but also on how this reconfiguration is implemented. Under ideal continuous activation, the PA position can be adjusted along the waveguide with arbitrary spatial resolution, providing a useful performance benchmark. Related studies on FA and MA systems have similarly examined the transition from continuous-position models to finite-port or discrete-position implementations~\cite{Psomas2023ContinuousFAS,Wu2025DiscreteMA}. Practical PA systems are likewise more likely to rely on a finite number of candidate positions rather than continuously adjustable radiation points.

Finite PA deployment introduces spatial quantization because the selected PA can only approximate the radiation position preferred by a user. The resulting position mismatch changes the PA--user distance and consequently affects the received signal strength. Characterizing the gap between discrete and continuous activation is therefore essential for determining the number of candidate PAs required to approach the continuous-activation benchmark.

To characterize the performance limits of PASS, Ouyang \emph{et al.} investigated its achievable array gain in~\cite{Ouyang2025ArrayGain}. Closed-form analytical results were derived to characterize the effects of antenna spacing and the number of antennas, revealing configurations that maximize the achievable array gain. Wang \emph{et al.} subsequently developed comprehensive hardware and signal models for PA systems in~\cite{Wang2025PASSBeamforming} and compared continuous and discrete activation strategies. Their numerical results showed that discrete activation can incur only a limited performance loss when sufficiently dense candidate positions are available. In addition, practical propagation effects within the dielectric waveguide have received increasing attention, and the impact of in-waveguide attenuation on PA system performance has been analyzed in \cite{Zhang2026PhysicalLayerSecurity}.

These studies suggest that discrete PA architectures can approach continuous-activation performance with a sufficiently dense deployment. Nevertheless, the continuous-to-discrete gap has mainly been evaluated through array-gain analysis or numerical comparison. The effects of finite PA deployment on the underlying PA--user distance statistics and the resulting OP and ER under different multiple-access schemes have not yet been fully characterized.

\subsection{Motivation and Contributions}

The discretization introduces spatial quantization, which changes the PA--user distance distribution and directly affects communication reliability and spectral efficiency. Therefore, it remains unclear how the number of discrete PAs influences system performance and how efficiently a practical DAPA system can approach the continuous-activation benchmark. Motivated by these issues, this paper investigates the performance of DAPA-assisted multiuser communications under both OMA and NOMA in terms of OP, ER, and DCGR.

The main contributions of this work are summarized as follows.

\begin{itemize}

\item
A multiuser communication system aided by DAPA is proposed, where a finite number of PAs are uniformly deployed along a dielectric waveguide and selectively activated according to user locations. By explicitly incorporating the system geometry and discrete antenna deployment, closed-form statistical characterizations of the distance between PAs and users are derived, establishing the analytical foundation for subsequent performance evaluation.

\item
Analytical expressions for the OP and ER are derived under both OMA and NOMA transmission schemes. The analysis explicitly captures the influence of discrete PA deployment, user ordering, and SIC decoding. Furthermore, two endpoint configurations, namely the single-PA case ($N=1$) and the continuous-activation limit ($N\to\infty$), are investigated as the fixed-PA reference and the continuous-activation reference, respectively. Under OMA, these two cases characterize the performance endpoints induced by nearest-PA selection, whereas under NOMA they serve as reference configurations for evaluating the user-dependent and system-level performance.

\item
A new performance metric termed DCGR is proposed to quantify how efficiently finite deployment of discrete PAs approaches the spatial gain achievable under ideal continuous activation. This metric provides an intuitive criterion for evaluating deployment efficiency and determining the minimum number of deployed PAs required to achieve performance close to that of continuous activation.

\item
Numerical simulations are conducted to validate the developed analytical framework and investigate the impact of key system parameters on system performance. The results demonstrate that DAPA systems can closely approach the continuous-activation reference using only a moderate number of candidate PAs under the considered system settings.
\end{itemize}

\section{System Model}\label{Sec_system_model}

\begin{figure}
    \centering
    \includegraphics[width=\linewidth]{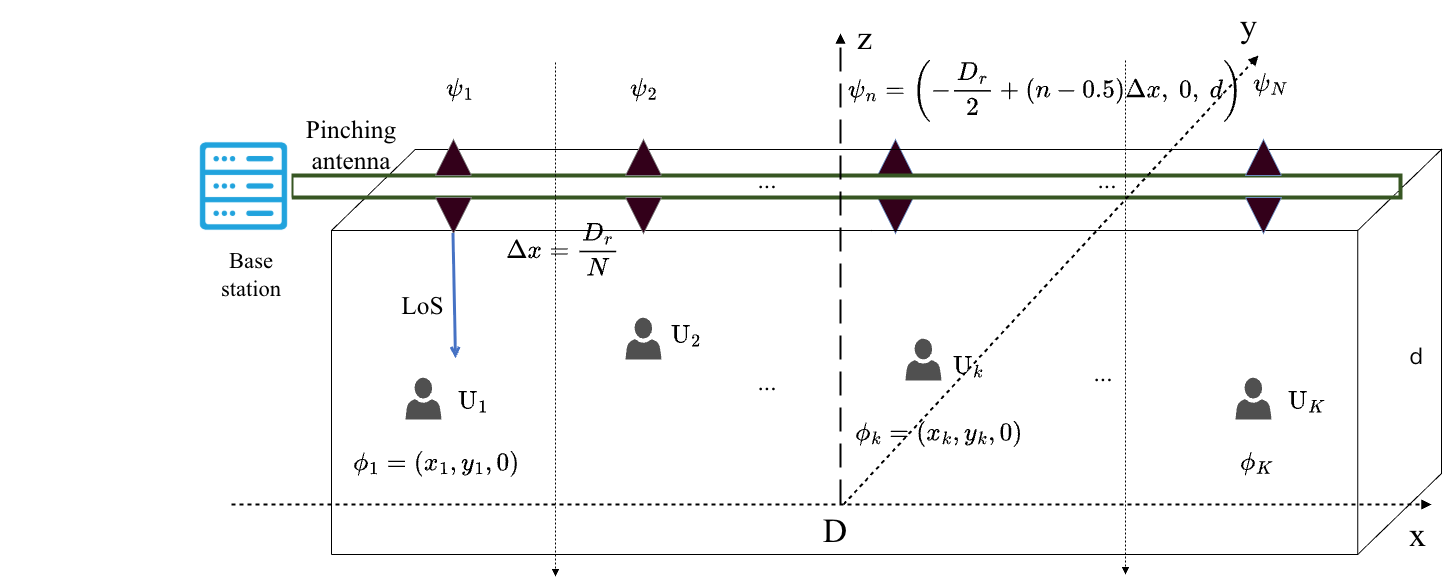}
    \caption{A DAPA system with $K$ users and $N$ PAs.}
    \label{fig_system_model}
\end{figure}

As illustrated in Fig.~\ref{fig_system_model}, we consider a DAPA system, where a dielectric waveguide is deployed parallel to the $x$-axis at a height $d$ above the ground plane. The waveguide supports low-loss surface-wave propagation and enables spatially distributed radiation through the selective activation of PAs. A total of $N$ PAs, denoted by $\mathrm{A}_n$, are uniformly deployed along the waveguide with inter-antenna spacing $\Delta x = \frac{D_r}{N}$. The coordinates of the $n$-th PA are given by
\begin{equation}\label{eq:PA_location}
    \begin{split}
        \psi_n = \left( -\frac{D_r}{2} + (n-0.5)\Delta x,\ 0,\ d \right),\quad n=1,2,\ldots,N.
    \end{split}
\end{equation}
We consider $K$ users, denoted by $\mathrm{U}_k$, which are uniformly distributed over a rectangular service region $\mathcal{A}$ of size $D_r \times D_l$ centered at the origin. For analytical tractability, all users are assumed to lie on a common horizontal plane, which is chosen as the reference plane $z=0$. Accordingly, the location of $\mathrm{U}_k$ is expressed as
\begin{equation}\label{eq:User_location}
    \begin{split}
        \phi_k = (x_k, y_k, 0),
    \end{split}
\end{equation}
where $x_k \in [-\frac{D_r}{2}, \frac{D_r}{2}]$ and $y_k \in [-\frac{D_l}{2}, \frac{D_l}{2}]$. Under this coordinate convention, $d$ represents the vertical separation between the dielectric waveguide and this plane.

\subsection{Channel Model}

The wireless channel between each PA and each user is mainly determined by the large-scale path loss associated with their propagation distance. Let $h_{n,k}$ denote the channel response between $\mathrm{A}_n$ and $\mathrm{U}_k$, which is modeled as
\begin{equation}\label{eq:Channel Model}
h_{n,k}=\sqrt{\frac{\eta}{Z_{n,k}}},
\end{equation}
where $\eta=\frac{c^2}{16\pi^2 f_c^2}$ is the path-loss-related constant, $c$ is the speed of light, and $f_c$ is the carrier frequency, $Z_{n,k} = \|\psi_n-\phi_k\|^2$ is the squared Euclidean distance between $\mathrm{A}_n$ and $\mathrm{U}_k$. This model explicitly captures the impact of the discrete PA deployment on the channel response, since different candidate
PAs lead to different values of $Z_{n,k}$.

\subsection{Signal Model of OMA}

\subsubsection{PA Selection Strategy}
We assume that TDMA is adopted as a representative OMA scheme, where each user is allocated an orthogonal time slot. Therefore, PA activation can be independently performed for each user without causing inter-user interference. This choice is particularly suitable for the considered DAPA architecture because the activated PA can be independently selected for each scheduled user in its own time slot. For $\mathrm{U}_k$, the nearest-PA selection rule is adopted, i.e., the selected PA index is given by
\begin{equation}
n_k^\star
= \arg\min_{n\in\{1,\ldots,N\}} Z_{n,k}.
\end{equation}
Since $h_{n,k}$ is a monotonically
decreasing function of $Z_{n,k}$, the above rule is equivalent to selecting the PA that provides the maximum channel response for $\mathrm{U}_k$. For notational convenience, we denote the distance between $\mathrm{U}_k$ and its selected PA by $Z_k \triangleq  Z_{n_k^\star,k}$, and the corresponding channel coefficient by $h_k \triangleq h_{n_k^\star,k} = \sqrt{\frac{\eta}{Z_k}}$.

\subsubsection{Signal Model}
Under OMA transmission, the received signal at $\mathrm{U}_k$ is given by
\begin{equation}
y_k = \sqrt{P}h_k s_k + n_k,
\end{equation}
where $s_k$ denotes the transmitted symbol intended for $\mathrm{U}_k$ with $\mathbb{E}\{|s_k|^2\}=1$, and $n_k$ is the additive white Gaussian noise (AWGN) with variance $\sigma^2$. The achievable rate of $\mathrm{U}_k$ under OMA is expressed as
\begin{equation}\label{eq:R_OMA}
R_k^{\mathrm{OMA}}=\frac{1}{K}\log_2\left(1+\frac{P|h_k|^2}{\sigma^2}\right),
\end{equation}
where the factor $\frac{1}{K}$ accounts for the equal resource allocation among the $K$ users.

\subsection{Signal Model of NOMA}
\subsubsection{PA Selection Strategy}
Unlike OMA, NOMA activates one common PA to simultaneously serve a group of scheduled users through power-domain superposition coding. We consider a NOMA cluster consisting of $K$ scheduled users, where user scheduling and cluster formation are assumed to have been completed prior to transmission. In the considered DAPA-assisted NOMA system, we randomly select a user, denoted by $\mathrm{U}_i$, as the designated strongest user by activating its nearest PA, i.e., the selected PA index is given by
\begin{equation}
n^\star = \arg\min_{n\in\{1,\ldots,N\}} Z_{n,i}.
\end{equation}
Accordingly, the activated PA is expressed as $\psi^{\star} = \psi_{n^\star}$. All $K$ users within the considered NOMA cluster are served by this activated PA. For notational convenience, we denote the squared distance between $\mathrm{U}_k$ and $\psi^{\star}$ by $Z_k^{\mathrm{NOMA}} \triangleq  Z_{n^\star,k}$, and the corresponding channel response by $h_k^\mathrm{NOMA} \triangleq h_{n^\star,k} = \sqrt{\frac{\eta}{Z_k^\mathrm{NOMA}}}$. We thus have
\begin{equation}
Y_k = |h_k^\mathrm{NOMA}|^2 = \frac{\eta}{Z_k^\mathrm{NOMA}}, \quad k=1,\ldots,K.
\end{equation}

The designated strongest user is denoted by $\mathrm{U}_{(K)}$. Excluding the designated user, the remaining $K-1$ users are ordered in ascending order according to their channel gains. Let $\mathrm{U}_{(r)}$ denote the user with the $r$-th smallest channel gain, where $r=1,\ldots,K-1$. Throughout this paper, the term ``weak user'' refers to a user with a relatively small effective channel gain. Their channel gains are given by
\begin{equation}
Y_{(k)} = |h_{(k)}^\mathrm{NOMA}|^2 = \frac{\eta}{Z_{(k)}^\mathrm{NOMA}},
\end{equation}
where $Z_{(k)}^{\mathrm{NOMA}}\triangleq Z_{n^\star,(k)}$ denotes the squared distance between $\mathrm{U}_{(k)}$ and the activated PA.

\begin{remark}\label{remark-strongest-user}
Note that $Y_{(1)} \leq Y_{(2)} \leq \cdots \leq Y_{(K-1)}$. However, $Y_{(K)}$ is not necessarily greater than $Y_{(r)}$ for $r=1,\ldots,K-1$. This is because the activated PA is selected as the nearest PA to $\mathrm{U}_{(K)}$, whereas the remaining users share the same activated PA and thus cannot independently optimize their PA-user distances. $\mathrm{U}_{(K)}$ is statistically more likely to have the largest channel gain. Nevertheless, it is not guaranteed to have the largest instantaneous channel gain in every realization.
\end{remark}

\subsubsection{Signal Model}
The transmitted superimposed signal is expressed as
\begin{equation}
x = \sum_{k=1}^{K} \sqrt{\alpha_k P}\,s_{(k)},
\end{equation}
where $P$ denotes the total transmit power, $\alpha_k$ is the power allocation coefficient associated with the $k$-th SIC decoding stage, satisfying $\alpha_1 \geq \alpha_2 \geq \cdots \geq \alpha_{K}>0$ and $\sum_{k=1}^{K}\alpha_k=1$, and $s_{(k)}$ denotes the information-bearing symbol intended for $\mathrm{U}_{(k)}$ with $\mathbb{E}\{|s_{(k)}|^2\}=1$.

Accordingly, the received signal at $\mathrm{U}_{(i)}$ is given by
\begin{equation}
y_{(i)} = h_{(i)}x+n_{(i)}, \quad i=1,\ldots,K, 
\end{equation}
where $n_{(i)}$ is the additive white Gaussian noise with variance $\sigma^2$. For $i=1,\ldots,K-1$, $h_{(i)}$ denotes the effective channel response of the $i$-th ordered weak user, while $h_{(K)}$ denotes the effective channel coefficient of the designated strongest user.

Following the conventional NOMA protocol, SIC is performed according to the ascending order of channel gains. Specifically, $\mathrm{U}_{(i)}$ decodes and removes the signals intended for $\mathrm{U}_{(r)}$, $r<i$, before decoding its own signal, while treating the signals intended for users with stronger channel conditions as interference.

For $r=1, \ldots, i$, the achievable rate at which $\mathrm{U}_{(i)}$ decodes the signal intended for the $r$-th ordered weak user is given by
\begin{equation}\label{eq:decoding}
R_{i\rightarrow r}=\log_2\left(1+\frac{|h_{(i)}|^2\alpha_r P}{|h_{(i)}|^2\sum_{j=r+1}^{K}\alpha_j P+\sigma^2}\right).
\end{equation}
Therefore, the achievable rate of the $r$-th ordered weak user is
\begin{equation}
R_{w,(r)}= \min_{i\in\{r,\ldots,K\}}R_{i\rightarrow r}, \quad r=1,\ldots,K-1.
\end{equation}

For $\mathrm{U}_K$, after cancelling all weak-user signals, its achievable rate is given by
\begin{equation}
R_{(K)}= \log_2\left(1+\frac{|h_{(K)}|^2\alpha_K P}{\sigma^2}\right).
\end{equation}

\section{OMA Case}\label{Sec_OMA}
In this section, we analyze the OP, ER, and DCGR of the DAPA-assisted OMA system.

\subsection{Distance Statistics}

To characterize the performance of the DAPA system, it is first necessary to derive the statistical distribution of the squared distance between a randomly located user and its associated activated PA. Since PA selection follows the nearest-PA selection rule, the resulting distance distribution is determined by the geometric partition induced by discrete antenna deployment. The following lemma establishes the PDF and CDF of this squared distance.

\begin{lemma}\label{lemma-z1}
Consider the user and PA locations defined in \eqref{eq:PA_location} and \eqref{eq:User_location}. Let $Z$ denote the squared distance between a generic user and its selected PA under the OMA nearest-PA selection rule. The PDF and CDF of $Z$ are given as follows:

\begin{equation}
f_Z(z)=
\begin{cases}
0, &\zeta<0,\\
\frac{\pi}{\Delta x D_l }, & 0<\zeta \leq \omega^2,\\
\frac{2}{\Delta xD_l}\beta_1, &\omega^2 < \zeta \leq \Omega^2,\\
\frac{2}{\Delta x D_l}\lambda_1, &\Omega^2 < \zeta \leq \omega^2+\Omega^2,\\
0, & \zeta>\omega^2+\Omega^2,
\end{cases}
\end{equation}
where ${\zeta}=z - d^2$, $\omega=\min\{\frac{\Delta x}{2}, \frac{D_l}{2}\}$, $\Omega=\max\{\frac{\Delta x}{2}, \frac{D_l}{2}\}$, $\beta_1=\arcsin{\frac{\omega}{\sqrt{\zeta}}}$, $\lambda_1=arcsin\frac{\omega}{\sqrt{\zeta}}-arcsin\sqrt{1-\frac{\Omega^2}{\zeta}}$,
and

\begin{equation}
F_Z(z)=
\begin{cases}
0, &\zeta<0,\\
\gamma_1, & 0\leq \zeta \leq \omega^2,\\
\epsilon_1, &\omega^2 < \zeta \leq \Omega^2,\\
\theta_1, &\Omega^2 < \zeta \leq \omega^2+\Omega^2,\\
1, & \zeta>\omega^2+\Omega^2,
\end{cases}
\end{equation}
where $\gamma_1=\frac{\pi\zeta}{\Delta xD_l}$, $\epsilon_1=\frac{2}{\Delta xD_l}(\omega\sqrt{\zeta-\omega^2}+\zeta\arcsin{\frac{\omega}{\sqrt{\zeta}}})$, $\theta_1=\frac{4}{\Delta xD_l}(\frac{\omega\sqrt{\zeta-\omega^2}+\Omega\sqrt{\zeta-\Omega^2}}{2}+\frac{\zeta}{2}(\arcsin{\frac{\omega}{\sqrt{\zeta}}}-\arcsin{\sqrt{1-\frac{\Omega^2}{\zeta}}}))$.
\end{lemma}

\subsection{Outage Probability}

The OP of $\mathrm{U}_k$ is defined as
\begin{equation}\label{Pin-U-OMA}
\begin{split}
\mathbb{P}_k^{\mathrm{OMA}}=\Pr\!\left( R_k^\mathrm{OMA}<\bar{R} \right),
\end{split}
\end{equation}
where $\bar{R}$ is the target rate. Based on the distance distribution established in Lemma~\ref{lemma-z1}, the OP under OMA can be derived by transforming the rate constraint into an equivalent distance threshold condition. The resulting expression is summarized in the following theorem.

\begin{theorem}\label{theorem-pin-oma}
In the DAPA-assisted OMA system, the OP of $\mathrm{U}_k$ is given by
\begin{equation}\label{Pin-U-OMA-OP}
\begin{split}
\mathbb{P}_k^\mathrm{OMA}=1-F_Z(a),
\end{split}
\end{equation}
where $a=\frac{\eta P}{(2^{K\bar{R}}-1)\sigma^2}$.
\end{theorem}
\begin{proof}

Based on \eqref{eq:R_OMA}, 
\begin{equation}
\begin{aligned}
\mathbb{P}_k^{\mathrm{OMA}}
&=\Pr\left(\frac{1}{K}\log_2\left(1+\frac{\eta P}{Z_k\sigma^2}\right)<\bar R\right) \\
&=\Pr\left(Z_k>\frac{\eta P}{(2^{K\bar R}-1)\sigma^2}\right) =1-F_Z(a),
\end{aligned}    
\end{equation}
where $F_Z(\cdot)$ is the CDF derived in Lemma~\ref{lemma-z1}. This completes the proof.
\end{proof}

To further reveal the fundamental reliability behavior of the proposed DAPA-assisted OMA system, it is instructive to examine the outage performance in the high-SNR regime. Such an asymptotic analysis provides insight into whether outage events can be completely eliminated and helps characterize the inherent limitations imposed by discrete PA activation. The corresponding high-SNR outage behavior is summarized in the following proposition.

\begin{proposition}\label{prop-oma-highSNR}
In the high-SNR regime, $\mathbb{P}^\mathrm{OMA}$ converges to zero, i.e.,
\begin{equation}
\begin{split}
\mathbb{P}^{\mathrm{OMA},\infty}=0.
\end{split}
\end{equation}
Specifically, $\mathbb{P}^\mathrm{OMA}$ becomes zero for $\rho \geq\frac{((\frac{\Delta x}{2})^2+(\frac{D_l}{2})^2+d^2)(2^{K\bar{R}}-1)}{\eta}$, where $\rho=\frac{P}{\sigma^2}$.

\begin{proof}
When $\rho \to \infty$, the argument of $F_Z(\cdot)$ goes to $+\infty$, and since $F_Z(\infty)=1$, it follows that $\mathbb{P}^{\mathrm{OMA}}=1-F_Z(\infty)=0$. This completes the proof.
\end{proof}
\end{proposition}

\subsubsection{Special Case of $N=1$}

\begin{corollary}\label{corollary-oma-n1}
When $N=1$ for OMA, the OP of $\mathrm{U}_k$ is given by
\begin{equation}\label{eq:OMA_OP_N1}
\mathbb{P}_k^{\mathrm{OMA},(1)} = 1-F_Z^{(1)}\left(\frac{\eta P}
{\left(2^{K\bar{R}}-1\right)\sigma^2}\right).
\end{equation}
Moreover, the OP becomes zero when
\begin{equation}\label{eq:rho_n1}
\rho\geq\frac{2^{K\bar{R}}-1}{\eta}\left(d^2+\frac{D_r^2}{4}+\frac{D_l^2}{4}\right).
\end{equation}
\end{corollary}

\begin{proof}
For $N=1$, the only PA is located at $(0,0,d)$, so that $Z_k^{(1)}=x_k^2+y_k^2+d^2$, with CDF $F_Z^{(1)}(\cdot)$. Thus, \eqref{eq:OMA_OP_N1} follows directly from Theorem~\ref{theorem-pin-oma} by setting $\Delta x=D_r$. Since $|x_k|\leq \frac{D_r}{2}$ and $|y_k|\leq \frac{D_l}{2}$, $Z_k^{(1)}\leq d^2+\frac{D_r^2}{4}+\frac{D_l^2}{4}$. Applying this maximum distance to the zero-outage condition yields~\eqref{eq:rho_n1}. This completes the proof.
\end{proof}

\subsubsection{Special Case of $N\to\infty$}

\begin{corollary}\label{corollary-oma-ninf}
When $N\to \infty$ for OMA, the OP of $\mathrm{U}_k$ is given by
\begin{equation}\label{eq:oma_op_ninf}
\mathbb{P}_k^{\mathrm{OMA},(\infty)}=1-F_Z^{(\infty)}\left(\frac{\eta P}
{\left(2^{K\bar{R}}-1\right)\sigma^2}\right).
\end{equation}
Moreover, the OP becomes zero when
\begin{equation}\label{eq:rho_ninf}
\rho\geq\frac{2^{K\bar{R}}-1}{\eta}\left(d^2+\frac{D_l^2}{4}\right).
\end{equation}
\end{corollary}

\begin{proof}
As $N\to\infty$, $\Delta x\to0$, and the selected PA approaches $(x_k,0,d)$. Hence, $Z_k^{(\infty)}=y_k^2+d^2$, with CDF $F_Z^{(\infty)}(\cdot)$, and \eqref{eq:oma_op_ninf} follows directly from Theorem~\ref{theorem-pin-oma}. Since $|y_k|\leq \frac{D_l}{2}$, $Z_k^{(\infty)}\leq d^2+\frac{D_l^2}{4}$. Applying this maximum distance to the zero-outage condition yields~\eqref{eq:rho_ninf}. This completes the proof.
\end{proof}

\begin{remark}
The cases $N=1$ and $N\to\infty$ serve as the fixed-antenna baseline and the ideal continuous-activation benchmark, respectively. For $N=1$, no spatial PA-selection gain is available. In contrast, as $N\to\infty$, the horizontal quantization error vanishes and the PA can be activated at the user's horizontal projection on the waveguide. These two cases characterize the endpoint configurations used to evaluate finite discrete PA deployments.
\end{remark}

\subsection{Ergodic Rate} 

The ER of $\mathrm{U}_k$ is defined as
\begin{equation}\label{eq:oma-er}
    \begin{split}
        R_k^{\mathrm{erg,OMA}}=\mathbb{E}\!\left[R_k^{\mathrm{OMA}}\right]
=\mathbb{E}\!\left(\frac{1}{K}\log_2\left(1+\frac{\eta P}{Z_k\sigma^2}\right)\right),
    \end{split}
\end{equation}
where the expectation is taken with respect to the random squared distance $Z_k$ between $\mathrm{U}_k$ and its selected PA, which is induced by the random user location.

Averaging the instantaneous OMA rate over the distance distribution in Lemma~\ref{lemma-z1} yields the following ER expression.

\begin{theorem}\label{theorem-oma-er}
Using the Chebyshev--Gauss quadrature approximation, the ER of $\mathrm{U}_k$ in the DAPA system is approximated as
\begin{equation}\label{eq:oma-er-final}
    \begin{split}
        R_k^{\mathrm{erg,OMA}}\approx\frac{1}{K\ln2\Delta x D_l}\left(I_1+2I_2+2I_3\right),
    \end{split}
\end{equation}
where $I_1$ is given by
\begin{equation}
\begin{aligned}
I_1 =
&\pi\Bigg[(d^2+\omega^2)\ln\left(1+\frac{\eta P}{\sigma^2(d^2+\omega^2)}\right)
\\
&-d^2\ln\left(1+\frac{\eta P}{\sigma^2 d^2}\right)+\frac{\eta P}{\sigma^2}\ln\left(\frac{\sigma^2(d^2+\omega^2)+\eta P}{\sigma^2 d^2+\eta P}\right)\Bigg],
\end{aligned}
\label{eq:I1}
\end{equation}
while $I_2$ and $I_3$ correspond to the integrals over the intermediate and far-distance regions, respectively.
Specifically, $I_2 =\int_{d^2+\omega^2}^{d^2+\Omega^2}\arcsin\!\left(\frac{\omega}{\sqrt{\zeta}}\right)\ln\!\left(1+\frac{\eta P}{z\sigma^2}\right)\mathrm{d}z$, $I_3=\int_{d^2+\Omega^2}^{d^2+\omega^2+\Omega^2}\lambda_1\ln\!\left(1+\frac{\eta P}{z\sigma^2}\right)\mathrm{d}z$.
Their Chebyshev--Gauss approximations are given by
\begin{align}
I_2 &\approx \frac{\pi(\Omega^2-\omega^2)}{2n}\sum_{i=1}^{n}\sin\theta_i \nonumber\\
&\quad\times \arcsin\!\left(\omega\sqrt{\frac{2}{(\Omega^2-\omega^2)\cos\theta_i+\omega^2+\Omega^2}}\right) \nonumber\\
&\quad\times \ln\!\left(1+\frac{\eta P/\sigma^2}{d^2+\frac{\omega^2+\Omega^2}{2}+\frac{\Omega^2-\omega^2}{2}\cos\theta_i}\right),
\label{eq:I2_Cheby}\\
I_3 &\approx \frac{\pi\omega^2}{2n}\sum_{i=1}^{n}\sin\theta_i\left(\arcsin\frac{\omega}{\sqrt{\zeta_i}}-\arcsin\sqrt{1-\frac{\Omega^2}{\zeta_i}}\right) \nonumber\\
&\quad\times \ln\!\left(1+\frac{\eta P/\sigma^2}{d^2+\zeta_i}\right),
\end{align}
where $\theta_i=\frac{(2i-1)\pi}{2n}$ and $\zeta_i=\Omega^2+\omega^2\cos^2(\frac{\theta_i}{2})$.

\end{theorem}

\begin{proof}
The ER of user $k$ under discrete PA activation is given by the expectation of the achievable rate with respect to the random distance $Z$ between the user and its associated PA. Using the distance distribution derived in Lemma~\ref{lemma-z1}, the ER can be expressed as
    \begin{equation}
        \begin{aligned}
        R_k^{\mathrm{erg,OMA}}&=\mathbb{E}\left(\frac{1}K{}\log_2 \left(1+\frac{\eta P}{Z\sigma^2}\right)\right)\\
        &=\int_0^{\infty}f_Z\left(z\right)\frac{1}{K}\log_2\left(1+\frac{\eta P}{z\sigma^2}\right)\mathrm{d}z,
        \end{aligned}
    \end{equation}
where $f_Z(z)$ denotes the probability density function of $Z$, and $Z$ denotes a generic copy of $Z_k$. Expressing the expectation in \eqref{eq:oma-er} in terms of the PDF $f_Z(z)$ and substituting the piecewise result in Lemma~\ref{lemma-z1} yield \eqref{eq:oma-er-final}. While $I_1$ admits the exact closed-form expression in \eqref{eq:I1}, $I_2$ and $I_3$ involve coupled inverse-trigonometric and logarithmic terms and are therefore approximated using the Chebyshev--Gauss quadrature.

Specifically, applying the variable substitution
\begin{equation}
    \begin{split}
        z=\frac{\Omega^2-\omega^2}{2}x+d^2+\frac{\omega^2+\Omega^2}{2},
    \end{split}
\end{equation}
the integral $I_2$ is transformed into the standard Chebyshev form and can be approximated as shown in \eqref{eq:I2_Cheby}. Similarly, the integral $I_3$ involves logarithmic terms and does not yield a closed-form solution. Introducing the auxiliary variable
\begin{equation}
    \begin{split}
        \zeta(x)=\Omega^2+\frac{\omega^2}{2}(x+1),
    \end{split}
\end{equation}
the Chebyshev--Gauss quadrature is again applied to approximate $I_3$, which leads to the expression given in Theorem~\ref{theorem-oma-er}. This completes the proof.
\end{proof}

\subsubsection{Special Case of $N=1$}

For $N=1$, the DAPA system reduces to a fixed-PA baseline without spatial selection. The only PA is located at $(0,0,d)$, and hence $Z_k^{(1)}=x_k^2+y_k^2+d^2$. The achievable rate under OMA is given by
\begin{equation}\label{eq:oma-1-er}
\begin{aligned}
R_k^{\mathrm{erg},N=1}=&\frac{1}{K D_r D_l}
\int_{-\frac{D_r}{2}}^{\frac{D_r}{2}}\int_{-\frac{D_l}{2}}^{\frac{D_l}{2}}\\
&\log_2\left(1+\frac{\eta P}{\sigma^2(x^2+y^2+d^2)}\right)\,\mathrm{d}y\,\mathrm{d}x.    
\end{aligned}
\end{equation}

Unlike finite-PA selection with $N>1$, the single-PA configuration cannot reduce the horizontal PA--user distance and therefore serves as the fixed-PA reference for the subsequent DCGR analysis.

\subsubsection{Special Case of $N\to\infty$}

As $N\to\infty$, the inter-PA spacing approaches zero and the selected PA approaches $(x_k,0,d)$. Thus, $Z_k^{(\infty)}=y_k^2+d^2$, and the achievable rate under OMA is
\begin{equation}
R_k^{\mathrm{erg},N\to\infty}=\frac{1}{K D_l}\int_{-\frac{D_l}{2}}^{\frac{D_l}{2}}\log_2\left(1+\frac{\eta P}{\sigma^2(y^2+d^2)}\right)\,\mathrm{d}y .
\end{equation}

Since $Z_k^{(\infty)}\geq d^2$, the continuous-activation ER is upper bounded by
\begin{equation}\label{eq:oma-wu-er}
R^{d,\infty} = \frac{1}{K}\log_2\left(1+\frac{\eta\rho}{d^2}\right).
\end{equation}
Moreover, the nearest-PA distance under any finite deployment is no smaller than that under continuous activation. Therefore, $R_k^{\mathrm{erg,OMA}}(N) \leq R_k^{\mathrm{erg,OMA}}(\infty) \leq R^{d,\infty}$. Hence, $R_k^{\mathrm{erg,OMA}}(\infty)$ serves as the continuous-activation benchmark for finite DAPA deployments,
whereas $R^{d,\infty}$ is an upper bound corresponding to the minimum possible PA--user distance $d$.

\subsection{DCGR of OMA}

Let $R_{\mathrm{sum}}^{\mathrm{erg,OMA}}(N)=\sum_{k=1}^{K}R_k^{\mathrm{erg,OMA}}(N)$ denote the system ergodic sum rate with $N$ candidate PAs. Using the single-PA configuration and continuous activation as the two reference points, respectively, the OMA DCGR is defined as
\begin{equation}
\label{eq:oma-DCGR-sum}
\mathrm{DCGR}^{\mathrm{OMA}}(N)=
\frac{R_{\mathrm{sum}}^{\mathrm{erg,OMA}}(N)
-R_{\mathrm{sum}}^{\mathrm{erg,OMA}}(1)}
{R_{\mathrm{sum}}^{\mathrm{erg,OMA}}(\infty)
-R_{\mathrm{sum}}^{\mathrm{erg,OMA}}(1)}.
\end{equation}
Under OMA, all users have the same ER because they share the same location distribution, transmit power, and resource allocation. Hence, $R_{\mathrm{sum}}^{\mathrm{erg,OMA}}(N)=K R_k^{\mathrm{erg,OMA}}(N)$, and \eqref{eq:oma-DCGR-sum} can equivalently be evaluated as
\begin{equation}
\label{eq:oma-DCGR-equivalent}
\mathrm{DCGR}^{\mathrm{OMA}}(N)=
\frac{R_k^{\mathrm{erg,OMA}}(N)
-R_k^{\mathrm{erg,OMA}}(1)}
{R_k^{\mathrm{erg,OMA}}(\infty)
-R_k^{\mathrm{erg,OMA}}(1)}.
\end{equation}
Under nearest-PA selection, $R_k^{\mathrm{erg,OMA}}(1)\leq R_k^{\mathrm{erg,OMA}}(N)\leq R_k^{\mathrm{erg,OMA}}(\infty)$. Therefore, $0\leq\mathrm{DCGR}^{\mathrm{OMA}}(N)\leq1$, with $\mathrm{DCGR}^{\mathrm{OMA}}(1)=0$ and $\mathrm{DCGR}^{\mathrm{OMA}}(\infty)=1$.

\begin{remark}
For fixed $N$, increasing $D_r$ enlarges the inter-PA spacing $\Delta x=\frac{D_r}{N}$ and therefore the horizontal quantization error under nearest-PA selection, which degrades the finite-PA ER. In contrast, since $Z_k^{(\infty)}=y_k^2+d^2$, the continuous-activation benchmark is independent of $D_r$ for fixed $D_l$ and $d$. Hence, a larger service-region length generally requires more candidate PAs to maintain a prescribed DCGR, as demonstrated in Section~\ref{Sec_Numerical_Results}.
\end{remark}

\section{NOMA CASE}\label{Sec_NOMA}

In this section, we analyze the OP, ER and DCGR of the DAPA-assisted NOMA system.

\subsection{Distance Statistics}

For the NOMA analysis, it is convenient to characterize the statistical distribution of the channel gain rather than the distance itself. Since the channel gain is a monotonic transformation of the PA--user squared distance, its distribution can be obtained from the corresponding distance statistics.

\begin{lemma}\label{lemma-z2}
Let $Y=\frac{\eta}{Z}$ denote the channel gain corresponding to the squared PA--user distance $Z$. Since $Y$ is a monotonically decreasing transformation of $Z$, its CDF is given by $F_Y(y)=1-F_Z\left(\frac{\eta}{y}\right)$.

Accordingly, the PDF and CDF of $Y$ are given by
\begin{equation}
\label{eq:pdf_Y}
f_Y(y) =
\begin{cases}
0, & \zeta'<0,\quad \zeta'>\omega^2+\Omega^2,
\\
\iota, & 0\leq\zeta'\leq\omega^2,
\\
\kappa\beta_2, & \omega^2<\zeta'\leq\Omega^2,
\\
\kappa\lambda_2, & \Omega^2<\zeta'\leq\omega^2+\Omega^2,
\end{cases}
\end{equation}
and
\begin{equation}
\label{eq:cdf_Y}
F_Y(y)=
\begin{cases}
1, & \zeta'<0,
\\
1-\gamma_2, & 0\leq\zeta'\leq\omega^2,
\\
1-\epsilon_2, & \omega^2<\zeta'\leq\Omega^2,
\\
1-\theta_2, & \Omega^2<\zeta'\leq\omega^2+\Omega^2,
\\
0, & \zeta'>\omega^2+\Omega^2,
\end{cases}
\end{equation}
respectively, where $\zeta'=\frac{\eta}{y}-d^2$, $\iota=\frac{\pi\eta}{\Delta x D_l y^2}$, $\kappa=\frac{2\eta}{\Delta x D_l y^2}$, $\beta_2=\arcsin\left(
\frac{\omega}{\sqrt{\zeta'}}\right)$, $\lambda_2=\arcsin\left(\frac{\omega}{\sqrt{\zeta'}}\right)-\arcsin\left(\sqrt{1-\frac{\Omega^2}{\zeta'}}\right)$, $\gamma_2=\frac{\pi\zeta'}{\Delta x D_l}$, $\epsilon_2=\frac{2}{\Delta x D_l}\left[\omega\sqrt{\zeta'-\omega^2}+\zeta'\arcsin\left(\frac{\omega}{\sqrt{\zeta'}}\right)\right]$, $\theta_2=\frac{4}{\Delta x D_l}\Bigg[\frac{\omega\sqrt{\zeta'-\omega^2}+\Omega\sqrt{\zeta'-\Omega^2}}{2}+\frac{\zeta'}{2}
\left(\arcsin\left(\frac{\omega}{\sqrt{\zeta'}}\right)-\arcsin\left(\sqrt{1-\frac{\Omega^2}{\zeta'}}\right)\right)\Bigg]$.

According to the PA selection rule in Section~\ref{Sec_system_model}, the activated PA is the nearest PA to the designated strongest user. Hence, $f_{Y_{(K)}}(y)=f_Y(y)$, $F_{Y_{(K)}}(y)=F_Y(y)$. Conditioned on $n^\star=n$, let $Z_{w\mid n}$ denote the squared distance between $\mathrm{A}_n$ and a generic unordered weak user, and let $F_{Z_w\mid n}(\cdot)$ and $f_{Z_w\mid n}(\cdot)$ denote its CDF and PDF, respectively. 

Since $Y_w=\eta/Z_{w\mid n}$, the corresponding conditional CDF and PDF are given by 
\begin{equation}
\begin{aligned}
F_{Y_w\mid n}(y)
&=1-F_{Z_w\mid n}\left(\frac{\eta}{y}\right),\\
f_{Y_w\mid n}(y)
&=\frac{\eta}{y^2}
f_{Z_w\mid n}\left(\frac{\eta}{y}\right),
\qquad y>0.
\end{aligned}
\end{equation}
The unconditional PDF and CDF of the $r$-th ordered weak-user channel gain are $f_{Y_{(r)}}(y)=\sum_{n=1}^{N}p_n f_{Y_{(r)}\mid n}(y)$, and$F_{Y_{(r)}}(y)=\sum_{n=1}^{N}p_n F_{Y_{(r)}\mid n}(y)$, where $f_{Y_{(r)}\mid n}(y)=\frac{(K-1)!}{(r-1)!(K-r-1)!}f_{Y_w\mid n}(y)\left[F_{Y_w\mid n}(y)\right]^{r-1}
\left[1-F_{Y_w\mid n}(y)\right]^{K-r-1}$, $F_{Y_{(r)}\mid n}(y)=\sum_{j=r}^{K-1}
\binom{K-1}{j}\left[F_{Y_w\mid n}(y)\right]^j\left[1-F_{Y_w\mid n}(y)\right]^{K-1-j}$, and $p_n=\Pr(n^\star=n)$.

\end{lemma}

\subsection{Outage Probability}

In downlink NOMA, a user can decode its own message only after successfully decoding and removing all messages preceding it in the SIC decoding order. Accordingly, the OP of $\mathrm{U}_{(r)}$ is calculated as $\mathbb{P}_{(r)}^{\mathrm{NOMA}} = 1-\Pr\left(R_{r\rightarrow 1}\geq\bar{R}_1,\ldots,R_{r\rightarrow r}\geq\bar{R}_r\right)$, where $r=1,\ldots,K$. In particular, $\mathrm{U}_{(K)}$ is the designated strongest user and performs all SIC stages before decoding its own message.

The OPs of the designated strongest user and the ordered weak users are given in the following theorem.

\begin{theorem}\label{theorem-pin-noma}
For $\mathrm{U}_{(K)}$, its OP is given by
\begin{equation}
\label{eq:noma_op_strong_user}
\mathbb{P}_{(K)}^{\mathrm{NOMA}}
=
\begin{cases}
0,&\displaystyle \lambda_K<\mu_1,
\\
F_Y\!\left(\lambda_K\right),&\displaystyle\mu_1\leq\lambda_K<\mu_4,
\\
1,&\displaystyle\mu_4\leq\lambda_K.
\end{cases}
\end{equation}
where $\mu_1=\frac{\eta}{d^2+\omega^2+\Omega^2}$, $\mu_2 = \frac{\eta}{d^2+\Omega^2}$, $\mu_3 = \frac{\eta}{d^2+\omega^2}$, $\mu_4 = \frac{\eta}{d^2}$.

For the $r$-th ordered weak user, $r=1,\ldots,K-1$, its OP is given by
\begin{equation}
\label{eq:noma_op_weak_user}
\mathbb{P}_{(r)}^{\mathrm{NOMA}}
=
\begin{cases}
0,
& \lambda_r<\dfrac{\eta}{Z_{\max}^{\mathrm{w}}},
\\
F_{Y_{(r)}}\!\left(\lambda_r\right), & \dfrac{\eta}{Z_{\max}^{\mathrm{w}}}\leq\lambda_r<\dfrac{\eta}{d^2},
\\
1,& \dfrac{\eta}{d^2}\leq\lambda_r,
\end{cases}
\end{equation}
where $Z_{\max}^{\mathrm{w}}\triangleq d^2+\left(D_r-\frac{D_r}{2N}\right)^2+\frac{D_l^2}{4}$ is the maximum squared distance between an ordered weak user and the activated PA, $\lambda_r=\max_{1\leq k\leq r}\xi_k$, $\lambda_K=\max_{1\leq k\leq K}\xi_k$, and $\xi_k=\frac{\left(2^{\bar{R}_k}-1\right)\sigma^2}{P\left(\alpha_k-\left(2^{\bar{R}_k}-1\right)\displaystyle\sum_{j=k+1}^{K}\alpha_j\right)}$, if $\alpha_k>\left(2^{\bar R_k}-1\right)\sum_{j=k+1}^{K}\alpha_j$, otherwise, $\xi_k=+\infty$.
\end{theorem}

\begin{proof}
From~\eqref{eq:decoding}, $R_{i\to k}\geq\bar{R}_k$ is equivalent to $Y_{(i)}\geq\xi_k$ whenever $\alpha_k>(2^{\bar{R}_k}-1)\sum_{j=k+1}^{K}\alpha_j$. 
Otherwise, decoding is infeasible and $\xi_k=+\infty$.

For the $r$-th ordered weak user, successful SIC requires decoding the messages intended for users $1,\ldots,r$. Hence,
\begin{equation}
\bigcap_{k=1}^{r}\{Y_{(r)}\geq\xi_k\}=\{Y_{(r)}\geq\lambda_r\},
\end{equation}
where $\lambda_r=\max_{1\leq k\leq r}\xi_k$. Thus, the OP is $F_{Y_{(r)}}(\lambda_r)$ within the support $\frac{\eta}{Z_{\max}^{w}}\leq Y_{(r)}\leq\mu_4$, which yields \eqref{eq:noma_op_weak_user}.

Similarly, $\mathrm{U}_{(K)}$ must decode all $K$ messages, and hence its success event is $\{Y_{(K)}\geq\lambda_K\}$, where $\lambda_K=\max_{1\leq k\leq K}\xi_k$.
Since $F_{Y_{(K)}}(y)=F_Y(y)$ and $\mu_1\leq Y_{(K)}\leq\mu_4$, accounting for the support yields~\eqref{eq:noma_op_strong_user}.
\end{proof}

\subsubsection{Special Case of $N=1$}

\begin{corollary}\label{corollary:noma_op_n1}
When $N=1$ for NOMA, the OP of $\mathrm{U}_{(K)}$ is given by
\begin{equation}
\label{eq:noma_op_strong_n1}
\mathbb{P}_{(K)}^{\mathrm{NOMA},N=1}=
\begin{cases}
0,& \displaystyle \lambda_K<\mu_1,
\\
F_Y^{(1)}\!\left(\lambda_K\right), &\displaystyle\mu_1\leq\lambda_K<\mu_4,
\\
1, &\displaystyle\mu_4\leq\lambda_K.
\end{cases}
\end{equation}

For the $r$-th ordered weak user, $r=1,\ldots,K-1$, its OP is given by
\begin{equation}
\label{eq:noma_op_weak_n1}
\mathbb{P}_{(r)}^{\mathrm{NOMA},N=1}=
\begin{cases}
0, &\displaystyle\lambda_r<\mu_1,
\\
\displaystyle F_{Y_{(r)}}^{(1)}(\lambda_r),&\displaystyle\mu_1\leq\lambda_r<\mu_4,
\\
1, &\displaystyle\mu_4\leq\lambda_r,
\end{cases}
\end{equation}
where $F_{Y_{(r)}}^{(1)}(y)=\sum_{j=r}^{K-1}\binom{K-1}{j}\left[F_Y^{(1)}(y)\right]^j\times\left[1-F_Y^{(1)}(y)\right]^{K-1-j}$.

\end{corollary}

\begin{proof}
For $N=1$, the only PA is located at the center of the waveguide and is always activated. Since the user locations are independent, the $K-1$ weak-user channel gains are mutually independent and share the same CDF $F_Y^{(1)}(\cdot)$. Moreover, $Z_{\max}^{w}=d^2+\frac{D_r^2}{4}+\frac{D_l^2}{4}$, which gives
$\frac{\eta}{Z_{\max}^{w}}=\mu_1$. Substituting these relations into \eqref{eq:noma_op_strong_user} and \eqref{eq:noma_op_weak_user} yields \eqref{eq:noma_op_strong_n1} and \eqref{eq:noma_op_weak_n1}, respectively. This completes the proof.
\end{proof}

\subsubsection{Special Case of $N\to\infty$}

Let $F_Y^{(\infty)}(\cdot)$ denote the CDF of the designated strongest-user channel gain in the continuous-activation limit. It is obtained from the previously derived $F_Y(\cdot)$ by letting $\Delta x\to0$, and is given by
\begin{equation}
\label{eq:cdf_strong_ninf}
F_Y^{(\infty)}(y) =
\begin{cases}
0,&\displaystyle y<\frac{\eta}{d^2+\dfrac{D_l^2}{4}},
\\
\displaystyle1-\frac{2}{D_l}\sqrt{\frac{\eta}{y}-d^2},&\displaystyle\frac{\eta}{d^2+\dfrac{D_l^2}{4}}\leq y<\mu_4,
\\
1,&\displaystyle\mu_4\leq y.
\end{cases}
\end{equation}

\begin{corollary}
\label{corollary:noma_op_ninf}
When $N\to\infty$ for NOMA, the OP of the designated strongest user is given by
\begin{equation}
\label{eq:noma_op_strong_ninf}
\mathbb{P}_{(K)}^{\mathrm{NOMA},N\to\infty}
=
\begin{cases}
0,&\displaystyle \lambda_K<\frac{\eta}{d^2+\dfrac{D_l^2}{4}},
\\
F_Y^{(\infty)}\!\left(\lambda_K\right),&\displaystyle \frac{\eta}{d^2+\dfrac{D_l^2}{4}}\leq\lambda_K<\mu_4,
\\
1,&\displaystyle\mu_4\leq\lambda_K.
\end{cases}
\end{equation}

For the $r$-th ordered weak user, $r=1,\ldots,K-1$, its OP is given by
\begin{equation}
\label{eq:noma_op_weak_ninf}
\mathbb{P}_{(r)}^{\mathrm{NOMA},N\to\infty}
=
\begin{cases}
0,&\displaystyle\lambda_r<\frac{\eta}{d^2+D_r^2+\dfrac{D_l^2}{4}},
\\
F_{Y_{(r)}}^{(\infty)}\!\left(\lambda_r\right),&\displaystyle\frac{\eta}{d^2+D_r^2+\dfrac{D_l^2}{4}}\leq\lambda_r<\mu_4,
\\
1,&\displaystyle \mu_4 \leq \lambda_r.
\end{cases}
\end{equation}
where $F_{Y_{(r)}}^{(\infty)}(\cdot)$ denotes the CDF of the $r$-th ordered channel gain among the $K-1$ weak users and is given by
\begin{equation}
\label{eq:cdf_weak_ordered_ninf}
\begin{aligned}
F_{Y_{(r)}}^{(\infty)}(y)
&=
\frac{1}{D_r}\int_{-\frac{D_r}{2}}^{\frac{D_r}{2}}\sum_{j=r}^{K-1}\binom{K-1}{j}
\\
&\quad\times\left[F_{Y_w\mid x}^{(\infty)}(y)\right]^j\left[1-F_{Y_w\mid x}^{(\infty)}(y)\right]^{K-1-j}\,\mathrm{d}x,
\end{aligned}
\end{equation}
where $F_{Y_w\mid x}^{(\infty)}(\cdot)$ is the conditional CDF of the channel gain of a generic unordered weak user, conditioned on the commonly activated PA being located at $(x,0,d)$.
\end{corollary}

\begin{proof}
See Appendix~\ref{app:proof_noma_op_ninf}.
\end{proof}

\subsection{Ergodic Rate}

In the considered downlink NOMA system, the designated strongest user determines the common activated PA, while the remaining $K-1$ users are ordered in ascending order of their channel gains.

\begin{theorem}\label{theorem:noma-er}
Since $\mathrm{U}_{(K)}$ is served by its nearest PA, its channel gain $Y_{(K)}$ follows the nearest-PA channel distribution given in Lemma~\ref{lemma-z2}. Therefore, the ER of $\mathrm{U}_{(K)}$ is expressed as
\begin{equation}\label{eq:Er-strong-definition}
R_{(K)}^{\mathrm{erg}}=\mathbb{E}\left[\log_2\!\left(1+\frac{\alpha_K P Y_{(K)}}{\sigma^2}\right)\right],
\end{equation}
where the expectation is taken with respect to the random channel gain $Y_{(K)}$ of the designated user.

Using the PDF in Lemma~\ref{lemma-z2}, the above expectation can be computed as
\begin{equation}\label{eq:Er_Us}
R_{(K)}^{\mathrm{erg}} = I_{K1} + I_{K2} + I_{K3},
\end{equation}
where $I_{K1}$ is obtained in closed form as
\begin{equation}
\begin{aligned}
I_{K1}
=&\frac{\pi\eta}{\Delta x D_l\ln 2}\Bigg[-\frac{\ln\!\left(1+\frac{\alpha_KP}{\sigma^2}y\right)}{y}
\\
&+\frac{\alpha_KP}{\sigma^2}\ln\!\left(\frac{\frac{\alpha_KP}{\sigma^2}y}{1+\frac{\alpha_KP}{\sigma^2}y}\right)\Bigg]_{\mu_3}^{\mu_4}.
\label{eq:IK1-closed}
\end{aligned}
\end{equation}
The remaining terms are approximated using the $Q$-point Chebyshev--Gauss quadrature as
\begin{equation}
\begin{aligned}
I_{K2}
&\approx{}\frac{\pi\eta(\mu_3-\mu_2)}{Q\Delta xD_l}\sum_{q=1}^{Q}\frac{\sin\theta_q}{y_{2,q}^{2}}\arcsin\!\left(\frac{\omega}{\sqrt{\zeta'(y_{2,q})}}\right)
\\
&\quad\times\log_2\!\left(1+\frac{\alpha_KP}{\sigma^2}y_{2,q}\right),
\label{eq:IK2-closed}
\end{aligned}
\end{equation}
\begin{equation}
\begin{aligned}
I_{K3}
&\approx\frac{\pi\eta(\mu_2-\mu_1)}{Q\Delta xD_l}\sum_{q=1}^{Q}\frac{\sin\theta_q}{y_{3,q}^{2}}
\\
&\times\Bigg[\arcsin\!\left(\frac{\omega}{\sqrt{\zeta'(y_{3,q})}}\right)-\arcsin\!\left(\sqrt{1-\frac{\Omega^2}{\zeta'(y_{3,q})}}\right)\Bigg]
\\
&\times\log_2\!\left(1+\frac{\alpha_KP}{\sigma^2}y_{3,q}\right),
\label{eq:IK3-closed}
\end{aligned}
\end{equation}

where $\theta_q=\frac{(2q-1)\pi}{2Q}$, $q=1,\ldots,Q$, $y_{2,q}=\frac{\mu_3-\mu_2}{2}\cos\theta_q+\frac{\mu_3+\mu_2}{2}$, $y_{3,q}=\frac{\mu_2-\mu_1}{2}\cos\theta_q+\frac{\mu_2+\mu_1}{2}$, $Q$ denotes the number of quadrature nodes, and $\zeta'(y)$ is given in Lemma~\ref{lemma-z2}.

For the $r$-th ordered weak-user message, $r=1,\ldots,K-1$, all users $U_{(i)}$ with $i=r,\ldots,K$ are required to decode this message during the SIC procedure. According to \eqref{eq:decoding}, the instantaneous achievable rate of the $r$-th ordered weak-user message is
\begin{equation}\label{eq:instantaneous-rate-rth-weak}
\begin{aligned}
R_{w,(r)}
&=\min_{i\in\{r,\ldots,K\}}R_{i\rightarrow r}
\\
&=\log_2\!\left(1+\frac{\alpha_r P\min\!\left\{Y_{(r)},Y_{(K)}\right\}}{P\min\!\left\{Y_{(r)},Y_{(K)}\right\}\displaystyle\sum_{j=r+1}^{K}\alpha_j+\sigma^2}\right).
\end{aligned}
\end{equation}

Consequently, the ER of the $r$-th ordered weak-user message is given by
\begin{equation}\label{eq:Er-rth-weak}
\begin{aligned}
&R_{(r)}^{\mathrm{erg}}=
\frac{\alpha_r P\sigma^2}{\ln 2}\sum_{n=1}^{N}p_n\int_{0}^{\infty}\frac{\left[1-F_{Y_{(r)}|n}(y)\right]\left[1-F_Y(y)\right]}{\left(\sigma^2+Py\displaystyle\sum_{j=r}^{K}\alpha_j\right)}
\\
&\times\frac{1}{\left(\sigma^2+Py\displaystyle\sum_{j=r+1}^{K}\alpha_j\right)}\mathrm{d}y
\approx\frac{\pi\mu_4\alpha_rP\sigma^2}{2Q\ln 2}\sum_{n=1}^{N}p_n\sum_{q=1}^{Q}\sin\theta_q
\\
&\times\frac{\left[1-F_{Y_{(r)}|n}(y_{r,q})\right]\left[1-F_Y(y_{r,q})\right]}{\left(\sigma^2+Py_{r,q}\displaystyle\sum_{j=r}^{K}\alpha_j\right)\left(\sigma^2+Py_{r,q}\displaystyle\sum_{j=r+1}^{K}\alpha_j\right)},
\end{aligned}
\end{equation}
where $y_{r,q}=\frac{\mu_4}{2}\left(1+\cos\theta_q\right)$.
\end{theorem}

\begin{proof}
See Appendix~\ref{appendix:noma-er-proof}.
\end{proof}

\subsubsection{Special Case of $N=1$}

\begin{corollary}\label{corollary-er-noma-N1}
For $N=1$, the system reduces to a conventional fixed-antenna NOMA system without PA selection. The ER of $U_{(K)}$ is
\begin{equation}\label{eq:Er-strong-N1}
R_{(K)}^{\mathrm{erg},N=1}=\frac{\alpha_KP}{\ln 2}\int_{0}^{\infty}\frac{1-F_Y^{(1)}(y)}{\sigma^2+\alpha_KPy}\,\mathrm{d}y.
\end{equation}

For the $r$-th ordered weak-user message, $r=1,\ldots,K-1$, the ER is given by
\begin{equation}\label{eq:Er-weak-N1}
\begin{aligned}
R_{(r)}^{\mathrm{erg},N=1}
=&
\frac{\alpha_rP\sigma^2}{\ln 2}\int_{0}^{\infty}\frac{\left[1-F_{Y_{(r)}}^{(1)}(y)\right]\left[1-F_Y^{(1)}(y)\right]}{\left(\sigma^2+Py\displaystyle\sum_{j=r}^{K}\alpha_j\right)}
\\
&\times\frac{1}{\left(\sigma^2+Py\displaystyle\sum_{j=r+1}^{K}\alpha_j\right)}\,\mathrm{d}y,
\end{aligned}
\end{equation}
where $F_Y^{(1)}(y)$ and $F_{Y_{(r)}}^{(1)}(y)$ are given in the preceding single-PA outage analysis.
\end{corollary}

\begin{proof}
For $N=1$, the only PA is located at $(0,0,d)$ and is always activated, so that $p_1=1$. The designated-user channel-gain CDF becomes $F_Y^{(1)}(y)$, while
$F_{Y_{(r)}\mid 1}(y)=F_{Y_{(r)}}^{(1)}(y)$. Applying the complementary-CDF representation to \eqref{eq:Er-strong-definition} yields \eqref{eq:Er-strong-N1}. Substituting $p_1=1$ and $F_{Y_{(r)}\mid 1}(y)=F_{Y_{(r)}}^{(1)}(y)$ into \eqref{eq:Er-rth-weak} yields \eqref{eq:Er-weak-N1}. This completes the proof.
\end{proof}

\subsubsection{Special Case of $N\to\infty$}

\begin{corollary}\label{corollary-er-noma-infty}

For $N\to\infty$, the ER of $\mathrm{U}_{(K)}$ is given by
\begin{equation}\label{eq:Er-strong-infty}
\begin{aligned}
R_{(K)}^{\mathrm{erg},\infty}
={}&
\int_{\frac{\eta}{d^2+\left(\frac{D_l}{2}\right)^2}}^{\mu_4}\frac{\eta}{D_l y^2\sqrt{\frac{\eta}{y}-d^2}}
\\
&\quad\times\log_2\!\left(1+\frac{\alpha_KP}{\sigma^2}y\right)\mathrm{d}y.
\end{aligned}
\end{equation}

Conditioned on the activated PA being located at $(x,0,d)$, the CDF of the $r$-th ordered weak-user channel gain, $r=1,\ldots,K-1$, is
\begin{equation}\label{eq:cdf-rth-weak-infty-cond}
\begin{aligned}
F_{Y_{(r)}|x}^{(\infty)}(y)
={}&
\sum_{\ell=r}^{K-1}\binom{K-1}{\ell}\left[F_{Y_w|x}^{(\infty)}(y)\right]^{\ell}
\\
&\times\left[1-F_{Y_w|x}^{(\infty)}(y)\right]^{K-1-\ell}.
\end{aligned}
\end{equation}

The ER of the $r$-th ordered weak-user message is given by
\begin{equation}\label{eq:Er-rth-weak-infty}
\begin{aligned}
R_{(r)}^{\mathrm{erg},\infty}
=
&\frac{\alpha_rP\sigma^2}{D_r\ln 2}\int_{-\frac{D_r}{2}}^{\frac{D_r}{2}}
\int_{0}^{\infty}\frac{\left[1-F_{Y_{(r)}|x}^{(\infty)}(y)\right]\left[1-F_Y^{(\infty)}(y)\right]}{\left(\sigma^2+Py\displaystyle\sum_{j=r}^{K}\alpha_j\right)}
\\
&\times
\frac{1}{\left(\sigma^2+Py\displaystyle\sum_{j=r+1}^{K}\alpha_j\right)}\,\mathrm{d}y\,\mathrm{d}x,
\end{aligned}
\end{equation}
where $F_Y^{(\infty)}(y)$ and $F_{Y_w|x}^{(\infty)}(y)$ are given in the preceding
continuous-activation outage analysis.
\end{corollary}

\begin{proof}
See Appendix~\ref{appendix-proof-er-noma-infty}.
\end{proof}

\subsection{DCGR of NOMA}

The DCGR of NOMA quantifies the normalized ergodic sum-rate gain achieved by a finite discrete PA deployment relative to the single-PA configuration, with continuous activation serving as the benchmark. For $N$ candidate PAs, it is defined as
\begin{equation}
\label{eq:noma-DCGR-sum}
\begin{aligned}
\mathrm{DCGR}^{\mathrm{NOMA}}(N)
&=
\frac{G^{\mathrm{NOMA}}(N)}{G^{\mathrm{NOMA}}(\infty)}
\\
&=
\frac{R_{\mathrm{sum}}^{\mathrm{erg,NOMA}}(N)-R_{\mathrm{sum}}^{\mathrm{erg,NOMA}}(1)}
{R_{\mathrm{sum}}^{\mathrm{erg,NOMA}}(\infty)-R_{\mathrm{sum}}^{\mathrm{erg,NOMA}}(1)},
\end{aligned}
\end{equation}
where $R_{\mathrm{sum}}^{\mathrm{erg,NOMA}}(N) =\sum_{k=1}^{K}R_{(k)}^{\mathrm{erg,NOMA}}(N)$ is the system ergodic sum rate with $N$ candidate PAs, and
$G^{\mathrm{NOMA}}(N) =R_{\mathrm{sum}}^{\mathrm{erg,NOMA}}(N)-R_{\mathrm{sum}}^{\mathrm{erg,NOMA}}(1)$ denotes the corresponding ergodic sum-rate gain relative to the single-PA reference. Here, $N=1$ represents the fixed-antenna baseline, whereas $N\to\infty$ represents the ideal continuous-activation benchmark.

Unlike the symmetric OMA case, the NOMA users generally have different ER because of channel ordering, power allocation, and SIC decoding. Therefore, the NOMA DCGR is evaluated directly using the system ergodic sum rate in~\eqref{eq:noma-DCGR-sum}. When the single-PA and continuous-activation configurations provide the lower and upper benchmarks, respectively, i.e., $R_{\mathrm{sum}}^{\mathrm{erg,NOMA}}(1)\leq R_{\mathrm{sum}}^{\mathrm{erg,NOMA}}(N)\leq R_{\mathrm{sum}}^{\mathrm{erg,NOMA}}(\infty)$, the DCGR satisfies $0 \leq \mathrm{DCGR}^{\mathrm{NOMA}}(N)\leq 1$. In particular, $\mathrm{DCGR}^{\mathrm{NOMA}}(1)=0$ and $\mathrm{DCGR}^{\mathrm{NOMA}}(\infty)=1$ correspond to the single-PA reference and ideal continuous-activation benchmark, respectively.

\section{Numerical Results}\label{Sec_Numerical_Results}
In this section, numerical results are presented to validate the theoretical results derived in the preceding sections and evaluate the performance of the proposed DAPA system under both OMA and NOMA. Unless otherwise specified, the simulation parameters are summarized in Table~\ref{tab:simulation_parameters}, and Monte Carlo simulations are conducted under the same parameter settings as those used for the theoretical evaluations. The numerical results first compare the theoretical and simulation results for the OP and ER, and then examine the impact of the number of candidate PAs on system performance. The DCGR is further evaluated to quantify how closely a finite DAPA deployment approaches the continuous-activation benchmark. In addition, the minimum number of candidate PAs required to achieve a prescribed DCGR is investigated for different service-region sizes. For the NOMA simulations, to avoid the additional complexity of power-allocation optimization, a fixed feasible power-allocation vector $\alpha=[0.6,0.25,0.15]$ is adopted.

\begin{table}
    \centering
    \caption{Parameter Settings}
    \begin{tabular}{|p{0.24\textwidth}|p{0.16\textwidth}|}
         \hline
         Side length of the room& $D_r=30m$ \\
         \hline
         Side width of the room& $D_l=20m$\\
         \hline
         Vertical separation between the waveguide and the user plane& $d=5m$\\
         \hline
         Carrier frequency& $f_{c}=28\ GHz$\\
         \hline
         Number of PAs& $N=1,2,3,4,5,10,\infty$\\
         \hline
         The power allocation coefficients for $\mathrm{U}_1$, $\mathrm{U}_2$, $\mathrm{U}_3$ (NOMA)& $\alpha_1=0.6$, $\alpha_2=0.25$ and $\alpha_3=0.15$\\
         \hline
         Target data rate& $\bar{R}=0.5\ Mbps$\\
         \hline
         Bandwidth& $B=10\ MHz$\\
         \hline
         Number of points for Chebyshev--Gauss quadratures& $n=1000$\\
         \hline
    \end{tabular}
    \label{tab:simulation_parameters}
\end{table}

\subsection{OPs of the DAPA with OMA and NOMA}

\begin{figure}
    \centering
    \includegraphics[width=0.8\linewidth]{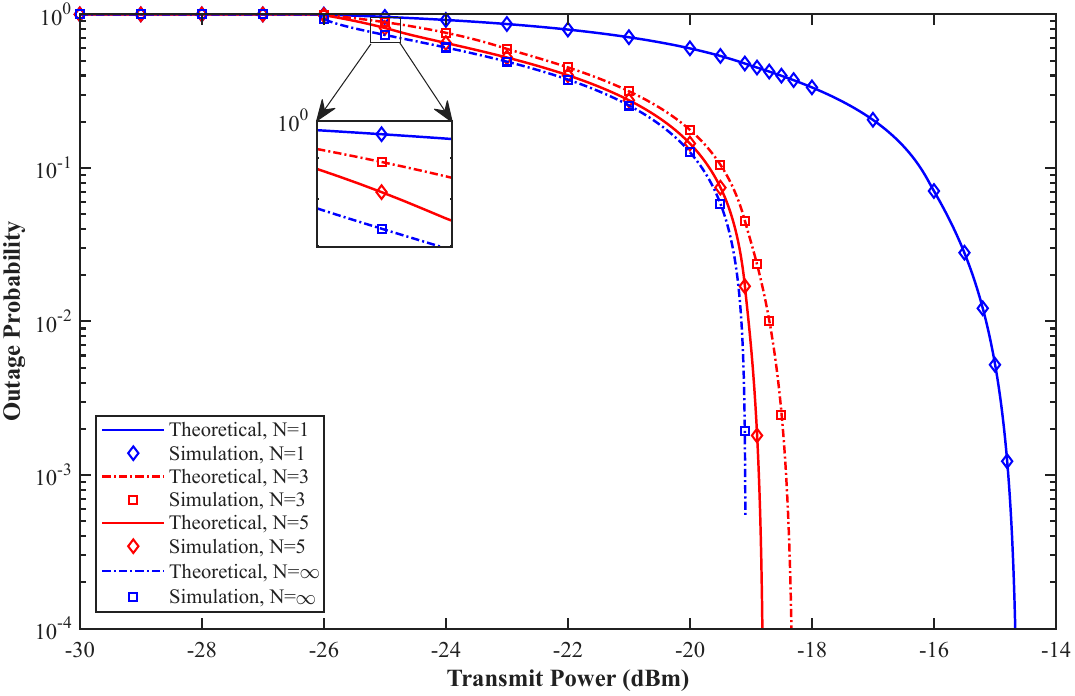}
    \caption{$\mathrm{U}_k$'s OPs in the DAPA with OMA with different $N$.}
    \label{fig-oma_n_Op}
\end{figure}

\begin{figure}
    \centering
    \includegraphics[width=0.8\linewidth]{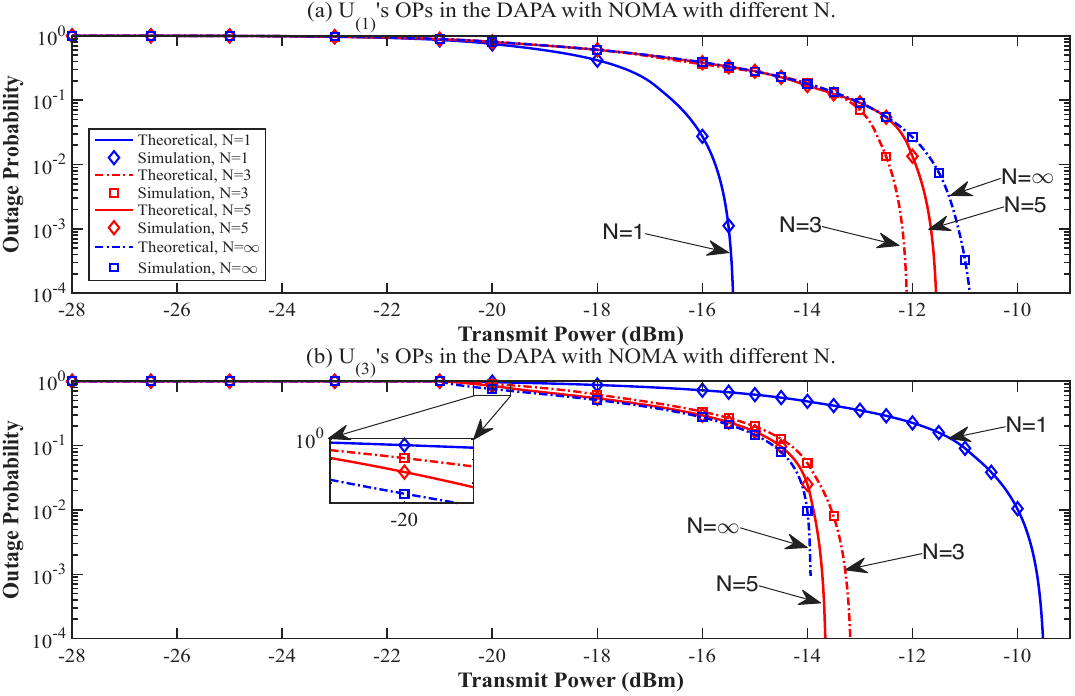}
    \caption{OPs in the DAPA with NOMA with different $N$.}
    \label{fig:noma-op}
\end{figure}

Fig.~\ref{fig-oma_n_Op} shows the OP of the DAPA-assisted OMA system for different numbers of candidate PAs. First, the simulation results closely match the corresponding analytical results in Theorem~\ref{theorem-pin-oma}, which validates the derived OP expression. Next, it is observed that the OP decreases as $N$ increases and eventually becomes zero at a finite transmit power. This improvement results from the nearest-PA selection strategy, since increasing $N$ reduces the horizontal quantization error and hence the PA--user distance, consistent with Lemma~\ref{lemma-z1} and the channel model in~\eqref{eq:Channel Model}. Furthermore, the performance gap between finite DAPA deployment and the continuous-activation benchmark gradually decreases with $N$, showing that a moderate number of candidate PAs can capture most of the outage-performance gain offered by continuous activation.

For the NOMA case, Fig.~\ref{fig:noma-op} compares the analytical and simulation OP results of $\mathrm{U}_{(1)}$ and $\mathrm{U}_{(3)}$ for different numbers of candidate PAs. First, the close agreement between the analytical and simulation results validates Theorem~\ref{theorem-pin-noma} and Corollaries~\ref{corollary:noma_op_n1} and~\ref{corollary:noma_op_ninf}. As shown in Fig.~\ref{fig:noma-op}(a), the OP of $\mathrm{U}_{(1)}$ increases with $N$. This is because the common PA is selected according to $\mathrm{U}_{(3)}$, and increasing $N$ does not necessarily reduce the PA--user distance for $\mathrm{U}_{(1)}$. In contrast, Fig.~\ref{fig:noma-op}(b) shows that the OP of $\mathrm{U}_{(3)}$ decreases as $N$ increases, since the activated PA becomes closer to $\mathrm{U}_{(3)}$ and thereby improves its channel condition. These opposite trends reveal the user-dependent impact of common-PA selection in NOMA. Moreover, the diminishing gap between finite $N$ and $N\to\infty$ for $\mathrm{U}_{(3)}$ further shows that a moderate number of candidate PAs can achieve most of the outage-performance gain of continuous activation.

\subsection{ER of the DAPA with OMA and NOMA}

\begin{figure}
    \centering
    \includegraphics[width=0.8\linewidth]{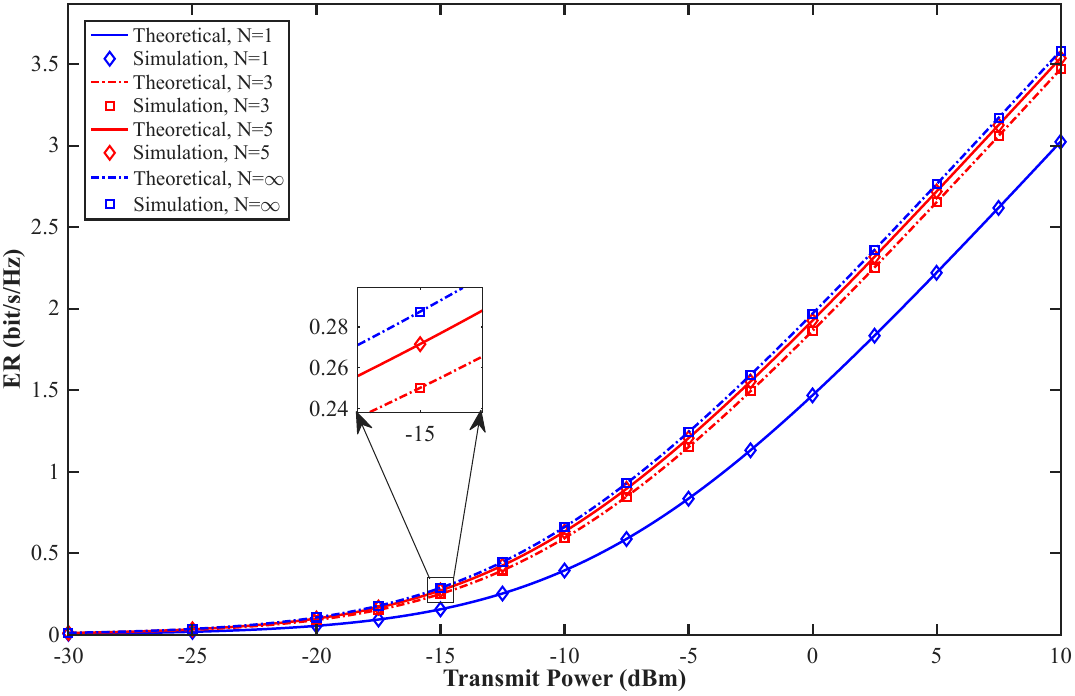}
    \caption{ER of $\mathrm{U}_k$ in the DAPA with OMA.}
    \label{fig-oma_Er}
\end{figure}

\begin{figure}
    \centering
    \includegraphics[width=0.8\linewidth]{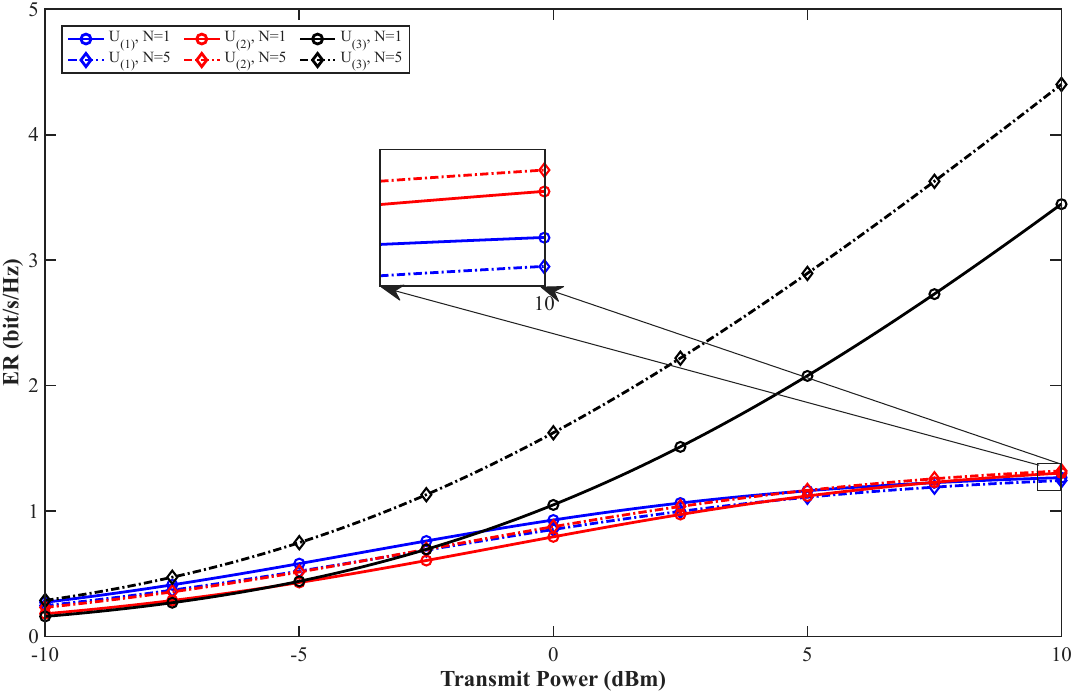}
    \caption{ER of $\mathrm{U}_k$ in the DAPA with NOMA.}
    \label{fig-noma_Er}
\end{figure}

Fig.~\ref{fig-oma_Er} shows the ER of the DAPA-assisted OMA system for different numbers of candidate PAs. First, the simulation results closely match the corresponding analytical results in~\eqref{eq:oma-er} and Theorem~\ref{theorem-oma-er}, which validates the derived ER expressions. Next, the ER increases with both the transmit power and $N$, since increasing the number of candidate PAs enables finer spatial selection and reduces the PA--user distance. However, the incremental ER gain gradually diminishes as $N$ increases. In particular, the small gap between $N=5$ and the continuous-activation benchmark shows that a moderate number of candidate PAs can achieve most of the ER gain offered by continuous activation.

Fig.~\ref{fig-noma_Er} compares the ERs of $\mathrm{U}_{(1)}$, $\mathrm{U}_{(2)}$, and $\mathrm{U}_{(3)}$ for $N=1$ and $N=5$. The close agreement between the analytical curves and simulation results validates Theorem~\ref{theorem:noma-er} and Corollary~\ref{corollary-er-noma-N1}. Increasing $N$ provides a pronounced ER improvement for $\mathrm{U}_{(3)}$, since the common PA is selected according to this user and a larger $N$ reduces its PA--user distance. In contrast, the effect on the ordered weak users is much smaller and user-dependent: the ER of $\mathrm{U}_{(1)}$ slightly decreases, whereas that of $\mathrm{U}_{(2)}$ slightly increases when $N$ increases from $1$ to $5$. This behavior results from the common-PA selection strategy, which directly improves the channel condition of $\mathrm{U}_{(3)}$ but does not simultaneously optimize those of the weak users. Moreover, as the transmit power increases, the ERs of $\mathrm{U}_{(1)}$ and $\mathrm{U}_{(2)}$ gradually approach the interference-limited ceilings determined by the fixed power-allocation coefficients, consistent with~\eqref{eq:instantaneous-rate-rth-weak}, whereas the ER of $\mathrm{U}_{(3)}$ continues to increase after cancelling the preceding streams through SIC. These results further reveal the user-dependent impact of common-PA selection in DAPA-assisted NOMA.

\subsection{DCGR of the DAPA with OMA and NOMA}

\begin{figure}
    \centering
    \includegraphics[width=0.8\linewidth]{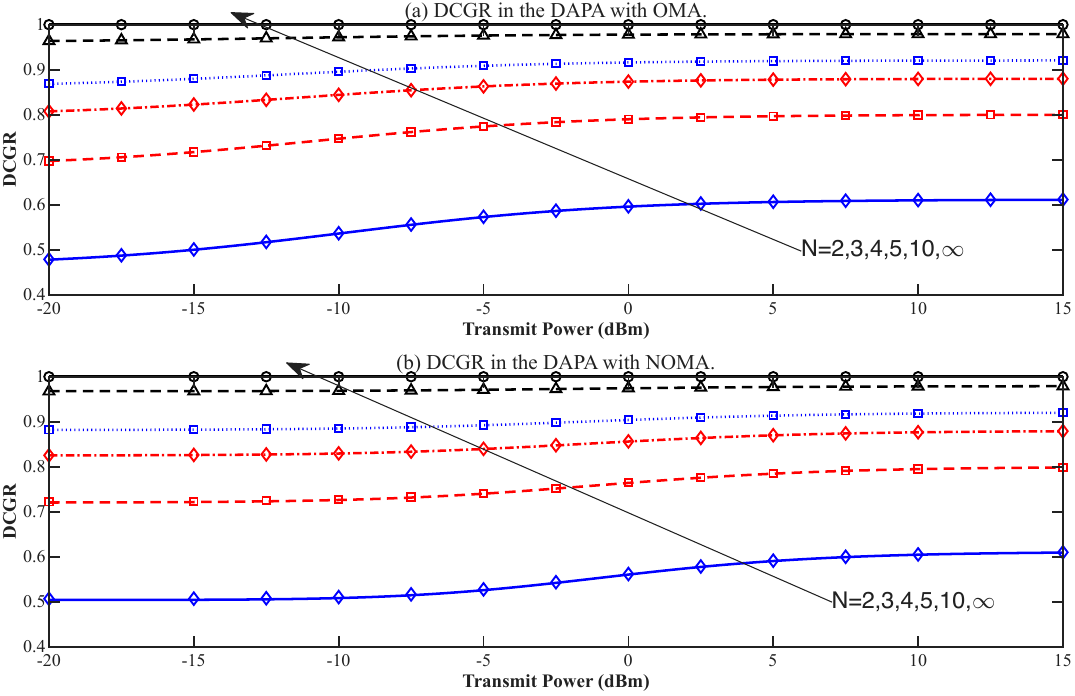}
    \caption{DCGR in the DAPA with different N.}
    \label{fig-oma_noma_DCGR}
\end{figure}

\begin{figure}
    \centering
    \includegraphics[width=0.8\linewidth]{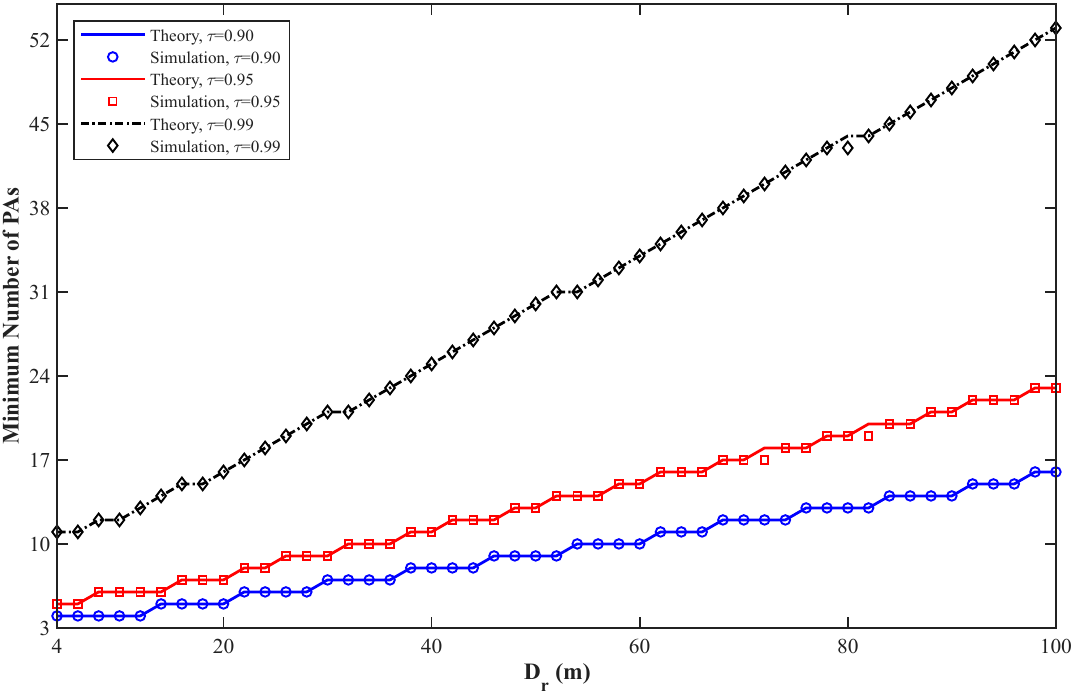}
    \caption{Minimum number of PAs for different room lengths under OMA.}
    \label{fig-oma-Nmin-Dr}
\end{figure}

Fig.~\ref{fig-oma_noma_DCGR}(a) illustrates the DCGR of the DAPA-assisted OMA system versus the transmit power for different numbers of candidate PAs. The analytical results closely match the simulations over the entire transmit-power range, validating the ER expressions and the DCGR formulation in~\eqref{eq:oma-DCGR-sum}. The DCGR increases with both $N$ and the transmit SNR, while the incremental gain diminishes as $N$ increases. For example, at a transmit power of $0$~dBm, the DCGR values are approximately $0.60$, $0.92$, and $0.98$ for $N=2$, $5$, and $10$, respectively. Thus, five candidate PAs already recover approximately $92\%$ of the ER gain offered by continuous activation, whereas further increasing $N$ provides only limited improvement.

Fig.~\ref{fig-oma_noma_DCGR}(b) shows the corresponding DCGR results for the DAPA-assisted NOMA system. The close agreement between the analytical and simulation results confirms the accuracy of the derived ER expressions and DCGR formulation. Similar to the OMA case, the DCGR increases with both $N$ and the transmit SNR, with diminishing improvement as $N$ increases. At a transmit power of $0$~dBm, the DCGR values are approximately $0.56$, $0.90$, and $0.98$ for $N=2$, $5$, and $10$, respectively. These results demonstrate that a moderate number of candidate PAs can recover most of the ER gain provided by the continuous-activation benchmark under both multiple-access schemes.

For a target DCGR $\tau$, let $N_{\min}^{\mathrm{OMA}}$ denote the minimum number of candidate PAs satisfying $\mathrm{DCGR}^{\mathrm{OMA}}(N)\geq\tau$.
Fig.~\ref{fig-oma-Nmin-Dr} shows $N_{\min}^{\mathrm{OMA}}$ versus the service-region length $D_r$. The staircase behavior results from the integer-valued nature of $N_{\min}^{\mathrm{OMA}}$. As $D_r$ increases from $4$ to $100$~m, $N_{\min}^{\mathrm{OMA}}$ increases from $4$ to $16$, from $5$ to $23$, and from $11$ to $53$ for $\tau=0.90$, $0.95$, and $0.99$, respectively. In particular,
the substantially larger PA requirement for $\tau=0.99$ indicates that recovering a larger fraction of the continuous-activation gain requires a denser PA deployment.
These results quantitatively confirm the trend discussed above.

\section{Conclusion}

In this paper, we investigated a DAPA system with OMA and NOMA transmissions by characterizing the distance statistics between the users and the activated PAs. The analytical and numerical results show that increasing the number of candidate PAs consistently improves the OMA performance and the performance of the designated NOMA user by reducing the corresponding PA--user distance. For the remaining NOMA users, however, the impact is user-dependent because all users share the PA selected according to $U_{(K)}$; consequently, their OP and ER do not necessarily improve with $N$. At the system level, the DCGR results under the considered parameter settings show that a moderate number of candidate PAs can recover most of the ER gain provided by the continuous-activation reference. Future work may extend the present analysis to more realistic channel environments, including fading, blockage, and hardware impairments. In addition, joint PA deployment, activation design, power allocation, and user scheduling can be further studied to support practical DAPA implementation in large wireless networks.

\appendices

\section{Proof of Corollary~\ref{corollary:noma_op_ninf}}
\label{app:proof_noma_op_ninf}

As $N\to\infty$, the activated PA approaches $(x_{(K)},0,d)$. Hence, the squared distance between the designated strongest user and the activated PA becomes $Z_{(K)}^{\mathrm{NOMA},\infty}=y_{(K)}^2+d^2$. Since $|y_{(K)}|\leq \frac{D_l}{2}$, $d^2\leq Z_{(K)}^{\mathrm{NOMA},\infty}\leq d^2+\frac{D_l^2}{4}$, and therefore we get
\begin{equation}
\frac{\eta}{d^2+\frac{D_l^2}{4}}\leq Y_{(K)}^{(\infty)}\leq\mu_4.
\end{equation}
Using this support together with \eqref{eq:cdf_strong_ninf} in the general OP expression \eqref{eq:noma_op_strong_user} yields \eqref{eq:noma_op_strong_ninf}.

For a generic weak user, the squared distance from the commonly activated PA is $Z_w^{(\infty)} = \left(x_w-x_{(K)}\right)^2+y_w^2+d^2$. Since $|x_w-x_{(K)}|\leq D_r$ and $|y_w|\leq \frac{D_l}{2}$, $d^2 \leq Z_w^{(\infty)} \leq d^2+D_r^2+\frac{D_l^2}{4}$, so that
\begin{equation}
\frac{\eta}{d^2+D_r^2+\frac{D_l^2}{4}} \leq Y_{(r)}^{(\infty)} \leq\mu_4.
\end{equation}

Conditioned on $x_{(K)}=x$, the $K-1$ weak-user locations are mutually independent and remain uniformly distributed over the service region. Hence, their channel gains are mutually independent and share the conditional CDF $F_{Y_w\mid x}^{(\infty)}(\cdot)$. The conditional CDF of the $r$-th ordered weak-user channel gain is therefore
\begin{equation}
\begin{aligned}
F_{Y_{(r)}\mid x}^{(\infty)}(y)
={}&
\sum_{j=r}^{K-1}
\binom{K-1}{j}
\left[F_{Y_w\mid x}^{(\infty)}(y)\right]^j
\\
&\times
\left[1-F_{Y_w\mid x}^{(\infty)}(y)\right]^{K-1-j}.
\end{aligned}
\end{equation}
Since $x_{(K)}$ is uniformly distributed over $[-\frac{D_r}{2},\frac{D_r}{2}]$, averaging the above conditional CDF over $x_{(K)}$ gives \eqref{eq:cdf_weak_ordered_ninf}. Combining this CDF and the above support with \eqref{eq:noma_op_weak_user} yields \eqref{eq:noma_op_weak_ninf}. This completes the proof.

\section{Proof of Theorem~\ref{theorem:noma-er}}
\label{appendix:noma-er-proof}

For the designated strongest user, substituting the piecewise PDF of $Y_{(K)}$ in Lemma~\ref{lemma-z2} into \eqref{eq:Er-strong-definition} decomposes the expectation over $[\mu_1,\mu_2]$, $[\mu_2,\mu_3]$, and $[\mu_3,\mu_4]$. Direct integration over $[\mu_3,\mu_4]$ yields \eqref{eq:IK1-closed}, while applying the corresponding affine transformations and the $Q$-point Chebyshev--Gauss quadrature to the other two intervals yields \eqref{eq:IK2-closed} and \eqref{eq:IK3-closed}. Therefore, \eqref{eq:Er_Us} follows.

We next consider the $r$-th ordered weak-user message. According to \eqref{eq:decoding}, the decoding SINR is monotonically increasing with the channel gain. Since $Y_{(r)}\leq Y_{(r+1)}\leq\cdots\leq Y_{(K-1)}$, the minimum decoding rate among the ordered weak users is attained at $\mathrm{U}_{(r)}$. However, $Y_{(K)}$ is not necessarily larger than $Y_{(r)}$. Hence, the effective channel gain is $\min\{Y_{(r)},Y_{(K)}\}$, which gives \eqref{eq:instantaneous-rate-rth-weak}.

Conditioned on $n^\star=n$, $Y_{(r)}$ depends on the $K-1$ weak-user locations, whereas $Y_{(K)}$ depends on the designated strongest user location. Since these locations are independent, $Y_{(r)}$ and $Y_{(K)}$ are conditionally independent. Moreover, all PA selection regions have the same geometry, so the conditional CDF of $Y_{(K)}$ is $F_Y(\cdot)$ for every $n$. Therefore,
\begin{align}
&\Pr\!\left(\min\{Y_{(r)},Y_{(K)}\}>y\,\middle|\,n^\star=n\right)
\notag\\
&\quad=\left[1-F_{Y_{(r)}\mid n}(y)\right]\left[1-F_Y(y)\right].
\label{eq:conditional-min-ccdf}
\end{align}

Let $\phi_r(y)$ denote the logarithmic rate function in \eqref{eq:instantaneous-rate-rth-weak}. Since $\phi_r(0)=0$, the complementary-CDF identity $\mathbb{E}[\phi_r(X)] =\int_0^\infty \phi_r'(y)\Pr(X>y)\,\mathrm{d}y$, together with \eqref{eq:conditional-min-ccdf} and averaging over the activated PA index with probabilities $p_n$, yields the exact integral expression in \eqref{eq:Er-rth-weak}.

Finally, both $Y_{(r)}$ and $Y_{(K)}$ are upper bounded by $\mu_4$, so the corresponding complementary CDF vanishes for $y\geq\mu_4$. Applying the affine transformation $y=\frac{\mu_4(1+t)}{2}$, $-1\leq t\leq1$, followed by the $Q$-point Chebyshev--Gauss quadrature with the nodes defined in the theorem yields the finite-sum approximation in \eqref{eq:Er-rth-weak}. This completes the proof.

\section{Proof of Corollary~\ref{corollary-er-noma-infty}}
\label{appendix-proof-er-noma-infty}

As $N\to\infty$, the activated PA approaches $(x_{(K)},0,d)$, and hence $Z_{(K)}^{\mathrm{NOMA},\infty}=y_{(K)}^2+d^2$. Differentiating the continuous-activation CDF in \eqref{eq:cdf_strong_ninf} gives
\begin{equation}\label{eq:pdf-strong-infty-proof}
f_{Y_{(K)}}^{(\infty)}(y)=\frac{\eta}{D_l y^2\sqrt{\frac{\eta}{y}-d^2}},
\end{equation}
for $\frac{\eta}{(d^2+D_l^2/4)}\leq y\leq\mu_4$. Substituting \eqref{eq:pdf-strong-infty-proof} into the expectation of the instantaneous rate of $\mathrm{U}_{(K)}$
yields \eqref{eq:Er-strong-infty}.

For the $r$-th ordered weak-user message, conditioned on $x_{(K)}=x$, the $K-1$ weak-user locations are mutually independent and remain uniformly distributed over the service region. Hence, their channel gains are mutually independent and share the conditional CDF $F_{Y_w\mid x}^{(\infty)}(\cdot)$, so standard order statistics give \eqref{eq:cdf-rth-weak-infty-cond}.

By the same monotonicity argument used in the proof of Theorem~\ref{theorem:noma-er}, the effective channel gain for the $r$-th ordered weak-user message is
$\min\{Y_{(r)},Y_{(K)}\}$. Conditioned on $x_{(K)}=x$, $Y_{(r)}$ depends on the weak-user locations, whereas $Y_{(K)}$ depends on $y_{(K)}$. Since the user coordinates are independent, these two channel gains are conditionally independent. Moreover, $y_{(K)}$ remains uniformly distributed over $[-\frac{D_l}{2},\frac{D_l}{2}]$ conditioned on $x_{(K)}=x$, and therefore the conditional CDF of $Y_{(K)}$ is $F_Y^{(\infty)}(\cdot)$. Thus,
\begin{align}
&\Pr\!\left(\min\{Y_{(r)},Y_{(K)}\}>y\,\middle|\,x_{(K)}=x\right)
\notag\\
&\quad=\left[1-F_{Y_{(r)}\mid x}^{(\infty)}(y)\right]\left[1-F_Y^{(\infty)}(y)\right].
\label{eq:ccdf-min-infty}
\end{align}

Following the complementary-CDF argument in the proof of Theorem~\ref{theorem:noma-er} and averaging over the uniformly distributed $x_{(K)}\in[-\frac{D_r}{2},\frac{D_r}{2}]$ yields \eqref{eq:Er-rth-weak-infty}. This completes the proof.

\bibliographystyle{IEEEtran}
\bibliography{ref}

\end{document}